\documentclass{aa}  
\usepackage{siunitx}
\usepackage{graphicx}	
\usepackage{amsmath}	
\usepackage{amssymb}
\usepackage{hyperref}
\usepackage{tabularx}
\usepackage{mathptmx}
\usepackage{txfonts}
\usepackage{academicons}
\usepackage[T1]{fontenc}
\usepackage{longtable}
\usepackage{subcaption}
\usepackage[symbol]{footmisc}
\usepackage{multicol, blindtext}
\usepackage{subcaption}
\usepackage[font=small]{caption}
\usepackage[normalem]{ulem}
\providecommand{\bjdtdb}{\ensuremath{\rm {BJD_{TDB}}}}

\providecommand{\msun}{\ensuremath{M_\odot}}
\providecommand{\rsun}{\ensuremath{R_\odot}}
\providecommand{\lsun}{\ensuremath{L_\odot}}
\providecommand{\msun}{\ensuremath{M}}
\providecommand{\rsun}{\ensuremath{R}}
\providecommand{\lsun}{\ensuremath{L}}

\providecommand{\fave}{\langle F \rangle} 
\providecommand{\fluxcgs}{10$^9$ erg s$^{-1}$ cm$^{-2}$}

\providecommand{\rst}{\ensuremath{\,{\rm R_\odot}}}

\providecommand{\arcsec}{$^{\prime \prime}$}
\providecommand{\arcmin}{$^{\prime}$}

\DeclareRobustCommand{\VAN}[3]{#2}
\let\VANthebibliography\thebibliography
\def\thebibliography{\DeclareRobustCommand{\VAN}[3]{##3}\VANthebibliography}
\usepackage{hyperref}
\usepackage{academicons}
\usepackage{xcolor}
\usepackage{dblfloatfix}
\usepackage{appendix}
\usepackage{dsfont}
\newcommand{\orcid}[1]{\href{https://orcid.org/#1}{\includegraphics[width=8pt]{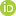}}}

\begin{document} 


   \definecolor{orcidlogocol}{HTML}{A6CE39}

\title{Discovery and characterization of a dense sub-Saturn TOI-6651b}
   \author{Sanjay Baliwal  \inst{1,2,*} \orcid{0000-0001-8998-3223} \and 
    Rishikesh Sharma \inst{1} \orcid{0000-0001-8983-5300} \and
    Abhijit Chakraborty \inst{1} \orcid{0000-0002-3815-8407} \and 
    Akanksha Khandelwal \inst{1} \orcid{0000-0003-0335-6435} \and
    K. J. Nikitha \inst{1} \orcid{0009-0000-4834-5612} \and
    Boris S. Safonov  \inst{3} \orcid{0000-0003-1713-3208} \and 
    Ivan A. Strakhov \inst{3} \orcid{0000-0003-0647-6133} \and 
    Marco Montalto \inst{4} \orcid{0000-0002-7618-8308} \and
    Jason D. Eastman\inst{5} \orcid{0000-0003-3773-5142} \and    
    David W. Latham\inst{5} \orcid{0000-0001-9911-7388} \and
    Allyson Bieryla\inst{5} \orcid{0000-0001-6637-5401} \and
    Neelam J.S.S.V. Prasad \inst{1} \orcid{0000-0003-0670-5821} \and 
    Kapil K. Bharadwaj  \inst{1} \orcid{0000-0003-1373-4583} \and 
    Kevikumar A. Lad  \inst{1} \and 
    Shubhendra N. Das  \inst{1,2} \orcid{0009-0006-9996-1814}\and     
    Ashirbad Nayak \inst{1} \orcid{0009-0001-3782-4308} }

   \institute{Astronomy $\&$ Astrophysics Division, Physical Research Laboratory, Ahmedabad 380009, India\\
   *\email{sanjaybaliwal1998@gmail.com}
    \and
    Indian Institute of Technology, Gandhinagar 382355, India
    \and
    Sternberg Astronomical Institute, Lomonosov Moscow State University, Moscow 119992, Russia
    \and
    INAF – Osservatorio Astrofisico di Catania, Via Santa Sofia 78, I-95123 Catania, Italy
    \and
    Center for Astrophysics | Harvard $\&$ Smithsonian, 60 Garden St., Cambridge MA 02138, USA
    }
    
   \date{Received ----; accepted ----}


\abstract
{We report the discovery and characterization of a transiting sub-Saturn exoplanet TOI-6651b using PARAS-2 spectroscopic observations. The host, TOI-6651 ($m_{V}\approx 10.2$), is a sub-giant, metal-rich G-type star with $[{\rm Fe/H}] = 0.225^{+0.044}_{-0.045}$, $T_{\rm eff} = 5940\pm110\ \mathrm{K}$, and $\log{g} = 4.087^{+0.035}_{-0.032}$. Joint fitting of the radial velocities from PARAS-2 spectrograph and transit photometric data from Transiting Exoplanet Survey Satellite (TESS) reveals a planetary mass of $61.0^{+7.6}_{-7.9}\ M_\oplus$ and radius of $5.09^{+0.27}_{-0.26}\ R_\oplus$, in a $5.056973^{+0.000016}_{-0.000018}$ day orbit with an eccentricity of $0.091^{+0.096}_{-0.062}$. TOI-6651b has a bulk density of $2.52^{+0.52}_{-0.44}\ \mathrm{g\ cm^{-3}}$, positioning it among the select few known dense sub-Saturns and making it notably the densest detected with TESS. TOI-6651b is consistent with the positive correlation between planet mass and the host star's metallicity. We find that a considerable portion $\approx$ 87\% of the planet's mass consists of dense materials such as rock and iron in the core, while the remaining mass comprises a low-density envelope of H/He. TOI-6651b lies at the edge of the Neptunian desert, which will be crucial for understanding the factors shaping the desert boundaries. The existence of TOI-6651b challenges conventional planet formation theories and could be a result of merging events or significant atmospheric mass loss through tidal heating, highlighting the complex interplay of dynamical processes and atmospheric evolution in the formation of massive dense sub-Saturns.}

\keywords{----}

\maketitle

\section{Introduction} \label{sec:intro}

\begin{figure*}[b!]
\centering
\includegraphics[width=0.85\columnwidth]{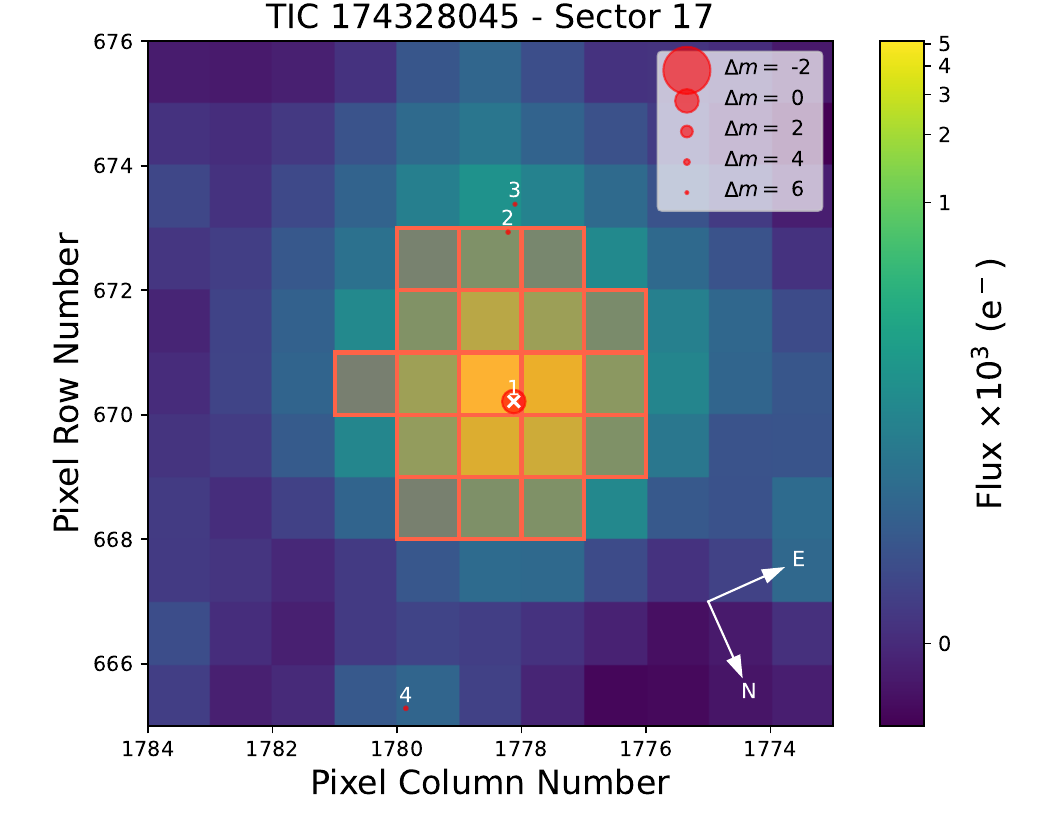}
\hspace{1.0cm}
\includegraphics[width=0.85\columnwidth]{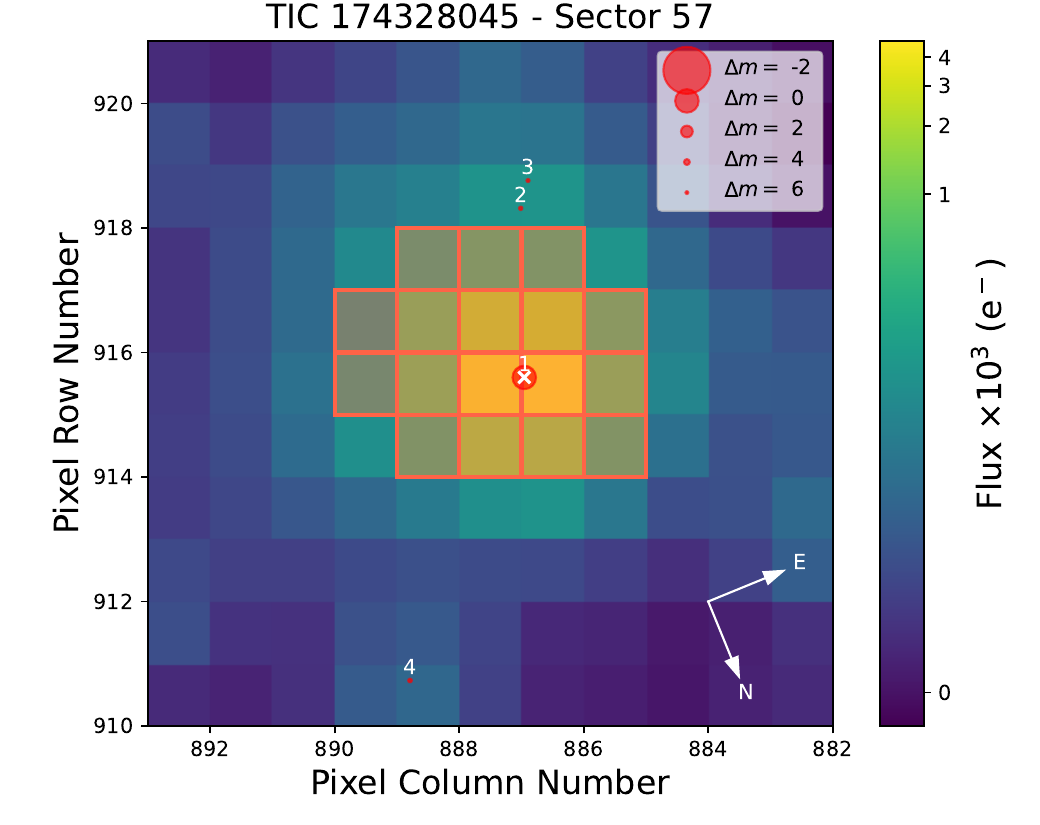}
    \caption{Target pixel file for TOI-6651 in sector 17 (on the left) and sector 57 (on the right) generated with \texttt{tpfplotter} \citep{tpfplot}. The squared region is the aperture mask used in the photometry, whereas the size of the individual red dot is the magnitude contrast ($\Delta m$) from TOI-6651. The position of TOI-6651 is marked with ``1''.}
\label{fig:tpfplot}
\end{figure*}

The sub-Saturn classification of exoplanets refers to planets larger than Neptune but smaller than Saturn, typically falling within the range of 4-8 $R_\oplus$ \citep{Petigura_2016}. Also known as Super-Neptunes, they are roughly ten times rarer than planets between the sizes of Earth and Neptune \citep{Brady_2018}. Sub-Saturns are often described as failed gas giants, possessing equally massive cores but significantly smaller total masses due to their much smaller accreted envelopes \citep{Millholland_2020}. These planets serve as valuable laboratories for studying the physics of envelope accretion, as Neptunes usually have envelope mass fractions ($f_{env}$) of less than 10\%, while gas giants are almost entirely composed of envelopes ($f_{env} \sim 100\%$) \citep{Petigura_2017}. The absence of these planets in our Solar System highlights the variety of planetary systems possible, and studying them around other stars offers useful insights into this diversity \citep{Petigura_2017, Fairnington2024}. Sub-Saturn planets show mass diversity regardless of size, spanning the mass range of 10-100 $M_\oplus$, strongly correlated to the host star metallicity \citep{Petigura_2017}.

The mass distribution between the core and envelope of sub-Saturns provides clues regarding their formation and evolution. A few theories have been proposed for the formation of sub-Saturns. One of them, proposed by \cite{Petigura_2017}, suggests the failure of runaway accretion, where the surrounding disk around a young star simply lacks sufficient material for the planet to become a full gas giant like Jupiter or Saturn. Alternatively, \cite{lee2016} and \cite{Lee_2018} proposed formation in a gas-depleted disk. In this scenario, the disk itself is depleted of gas, possibly due to stellar radiation or winds, leaving a sub-Saturn with limited building materials to reach the size of Jupiter or Saturn. Another proposed mechanism is formation through atmospheric mass loss, as suggested by \citep{Hallatt2022}. This theory proposes that sub-Saturns initially form as giant planets but lose their atmosphere over time, causing them to shrink down \citep{Owen2018}. Two main mechanisms are proposed for atmospheric mass loss: photoevaporation (e.g., \cite{Owen2018}; \cite{Owen_2019}; \cite{Owen_2012}; \cite{Jin_2014}; \cite{Lopez2017}) and high-eccentricity orbital migration followed by tidal interactions with the star, also known as tidal stripping (e.g., \cite{Matsakos_2016}; \cite{Owen_2018}). The formation pathways followed by a planet in its life among these two mechanisms result in diverse densities and masses.

So far\footnote{According to the NASA Exoplanet Archive \citep{NASA_EXO_Archive_Akeson_2013} as of July 17, 2024}, only 74 sub-Saturns are known with mass and radius constraints better than 50\% and 20\%, respectively, and 41 of them have orbital periods of less than 10 days (hot sub-Saturns). The noticeable scarcity of sub-Jovian planets on short orbits ($P<10$ days) is known as the ``sub-Jovian desert'' or ``hot Neptunian desert'' and that has been extensively discussed in the literature \citep[e.g., ][]{Lecavelier2007L,davis2009,szabo2011,Beauge2013,Mazeh2016}. The dearth is not due to observational biases, as close-in super-Neptunes or sub-Saturns are readily detectable through both transit and Radial Velocity (RV) methods. The origin of the Neptunian desert is also associated with the formation and evolution mechanism. For example, the upper edge of the Neptunian desert (see Figure~\ref{fig:neptunedesert}) is linked with the tidal heating mechanism as it is argued that denser and more massive sub-Saturns tend to lose their atmospheres primarily through tidal heating rather than photoevaporation. Photoevaporation remains the dominant mechanism for atmospheric loss in less massive and less dense sub-Saturns \citep{Vissapragada2022, Petigura_2017}. \citet{Thorngren2023} offer a different viewpoint suggesting that photoevaporation might not be entirely negligible even for massive sub-Saturns, implying a possible interplay between photoevaporation and tidal heating. Alternatively, a deeper understanding can also be gained by detailed estimation of the planet's internal structure and composition. This approach is particularly valuable because the formation pathway of these planets is closely linked to their core mass and the $f_{env}$ \citep{Fortney_2007,Petigura_2016}. Despite our current knowledge, gaps regarding the formation and evolution of these planets and their absence in our own solar system make them even more intriguing planets for further study.

Here, we present the discovery and characterization of TOI-6651b, a dense hot sub-Saturn transiting a metal-rich G-type star, utilizing data obtained from PARAS-2 and TESS. PARAS-2 is a high-resolution spectrograph (R$\approx$ 107,000) attached to the recently installed Physical Research Laboratory (PRL) 2.5m telescope at Mount Abu Observatory, located at Gurushikhar, Mount Abu, India \citep{paras2_design, prl2.5m_and_paras2}. In addition to TESS and PARAS-2, we have utilized speckle imaging data in this work, which is explained in Section~\ref{sec:obs}. In Section ~\ref{sec:analysis}, our focus is directed toward the host star and the comprehensive characterization of the star-planet system, achieved through the combined analysis of transit photometry and RV data. Section~\ref{sec:discussion} is dedicated to exploring the implications of our findings, while Section~\ref{sec:summary} provides a summary of the paper.


\section{Observations} \label{sec:obs}

\subsection{TESS photometry}

\begin{figure*}[t!]
\centering
\includegraphics[width=0.85\paperwidth]{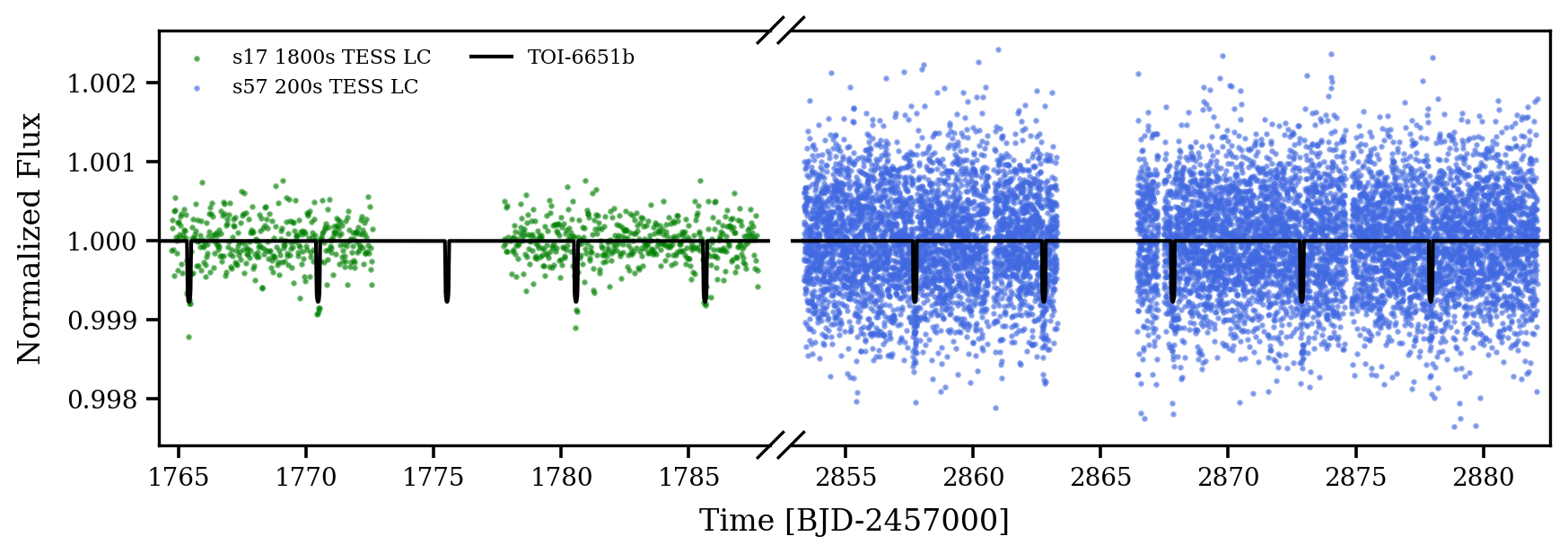}
\includegraphics[width=0.70\textwidth]{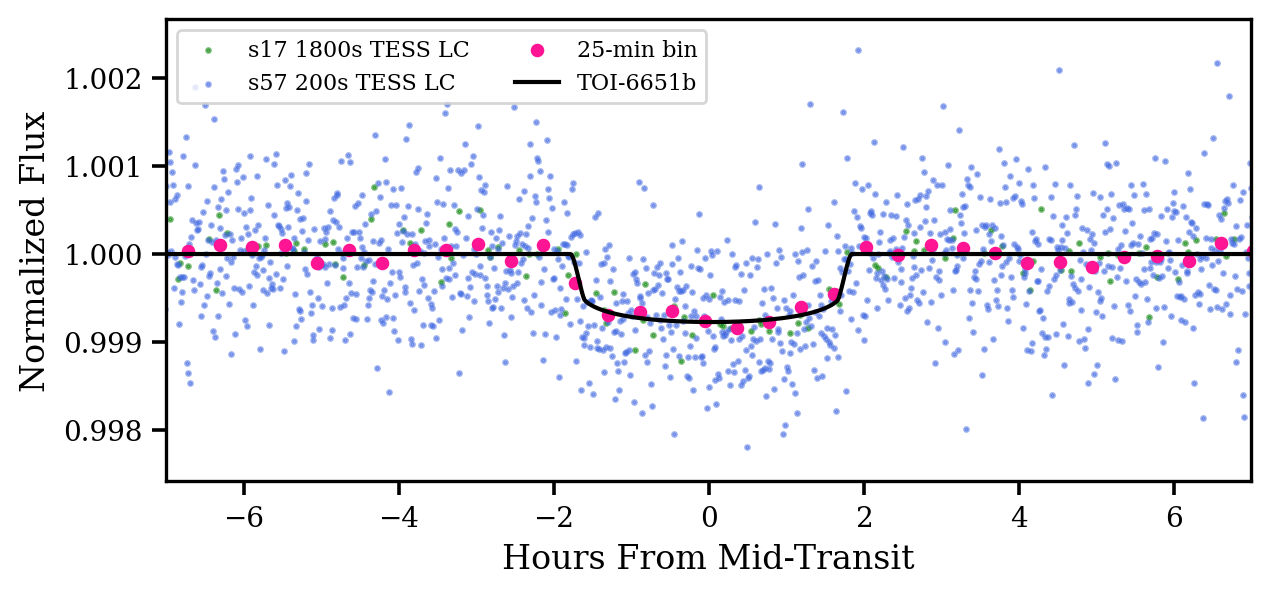}
    \caption{Detrended TESS light curve (LC) from sector 17 (1800-second cadence) and 57 (200-second cadence) shown in green and blue points, respectively. The upper panel displays the full TESS LC plotted against time, while the lower panel presents the phase-folded LC with pink dots representing 25-minute binned data points. The black solid line in both panels represents the best-fit transit model for TOI-6651b from our joint fit analysis (see Section~\ref{sec:global_modeling}).}
    \label{fig:tesslc}
\end{figure*}

TOI-6651 (TIC 174328045) was observed with TESS at a 30-minute cadence in sector 17 from October 8 to November 2, 2019, and at a 200-second cadence in sector 57 from September 30 to October 29, 2022. Additionally, TOI-6651 is slated for observation in sector 84 from October 1 to October 26, 2024, as indicated by the TESS-point Web Tool\footnote{\url{https://heasarc.gsfc.nasa.gov/wsgi-scripts/TESS/TESS-point_Web_Tool/TESS-point_Web_Tool/wtv_v2.0.py/}}. TOI-6651 was initially identified as a community TESS object of interest (CTOI; \cite{EXOFOP_Guerrero_2021}) by \cite{montalto22} using the \texttt{DIAmante}\footnote{\url{https://archive.stsci.edu/hlsp/diamante}} pipeline, which detected transit signals in light curve (LC) derived from TESS Full Frame Images (FFIs) of sector 17. Its status was subsequently elevated to TOI in September 2023 after LC generated from FFIs of sector 57 were scrutinized for transit signals by the Science Processing Operations Center (SPOC; \cite{spoc}) and Quick Look Pipeline (QLP; \cite{qlp1, qlp2}).

For photometric analysis, we used the Pre-search Data Conditioning Simple Aperture Photometry (PDCSAP) flux time series processed by the SPOC pipeline \citep{spoc}, publicly accessible through the Mikulski Archive for Space Telescopes (MAST\footnote{\url{https://mast.stsci.edu/portal/Mashup/Clients/Mast/Portal.html}}). 
For detrending and normalizing the LCs, we employed a Python package \texttt{citlalicue} \citep{citlalicue}, which involves Gaussian Process regression to remove systematics and stellar variability. The resulting LCs, post-procedure, are illustrated in Figure \ref{fig:tesslc}. We detected a total of 9 transits: 4 in sector 17 and 5 in sector 57. However, due to telescope reorientation during data transmission, one transit in sector 17 was omitted.

We also use \texttt{tpfplotter} \citep{tpfplot} to overplot the \textit{Gaia} DR3 \citep{gaia2023} sources onto the TESS target pixel file (TPF). This enabled us to identify potential sources of dilution in the TESS photometry, with a magnitude difference ($\Delta m$) limit of up to 6. The tpfplotter generated images are shown in Figure \ref{fig:tpfplot}. We found one additional source at the edge of the aperture mask of the target in sector 17. However, given its approximately six-magnitude difference compared to TOI-6651, this source is too faint to yield any significant dilution. No additional sources contributing to dilution were found in the observations of sector 57. Additionally, the pipeline analysis revealed that 99.63$\%$ of the light in the optimal aperture originates from the target itself rather than other stellar sources, indicating negligible dilution from faint background stars.

\subsection{Speckle observations}

\cite{labeyrie1970} introduced the concept of speckle interferometric imaging as a technique to minimize the effects of atmospheric turbulence that primarily limits the resolution of ground-based telescopes. This method involves acquiring multiple short-exposure images during which the atmosphere is partially frozen, which are then processed to reconstruct a single diffraction-limited image. Speckle imaging serves as a valuable follow-up observation to exoplanet detection studies by providing such high angular resolution images as it aids in disentangling signals from close companions to the host star, thus reducing the likelihood of false detections or incorrect estimation of stellar and planetary parameters \citep[e.g.,][]{2015ciardi, Hirsch2017, furlan2017}. Next, we present a discussion on the speckle imaging analysis of TOI-6651 using two different speckle cameras.

\subsubsection{Speckle imaging with PRL 2.5m telescope}

\begin{figure*}[t!]
    \centering
    \includegraphics[width=0.95\columnwidth]{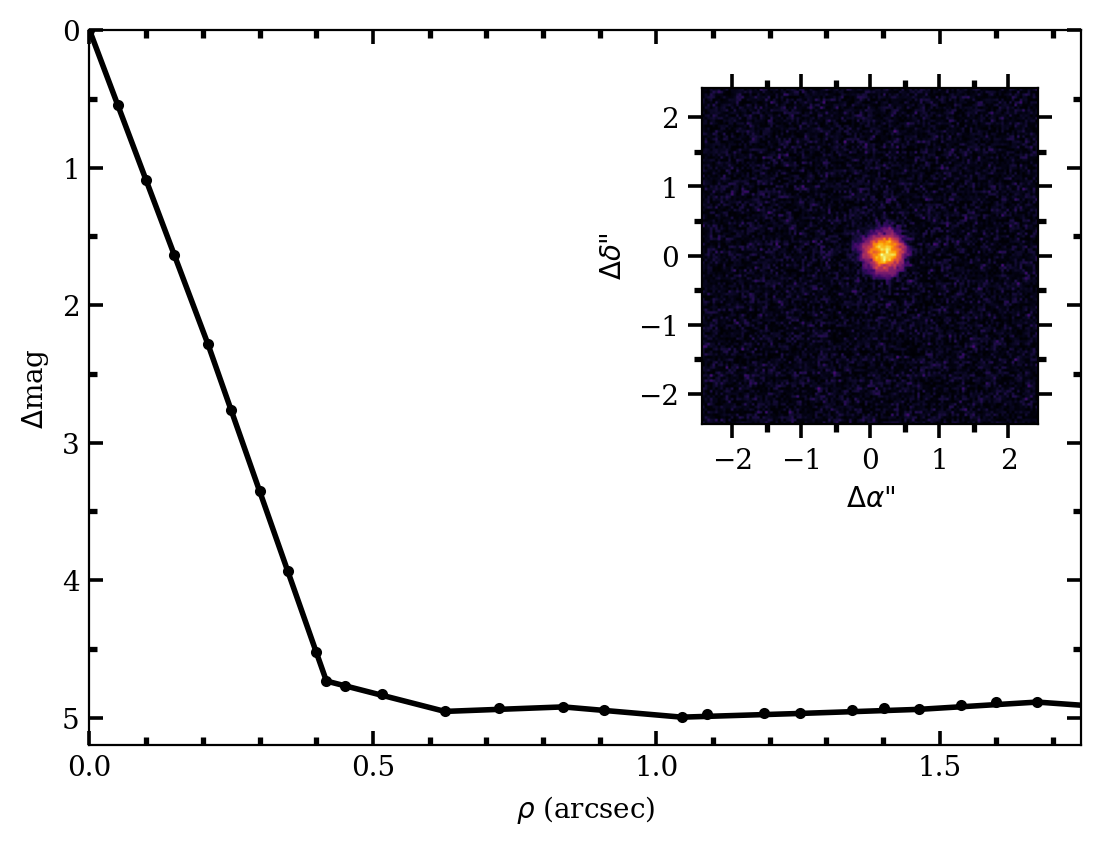}
    \hspace{0.7cm}
    \includegraphics[width=0.90\columnwidth]{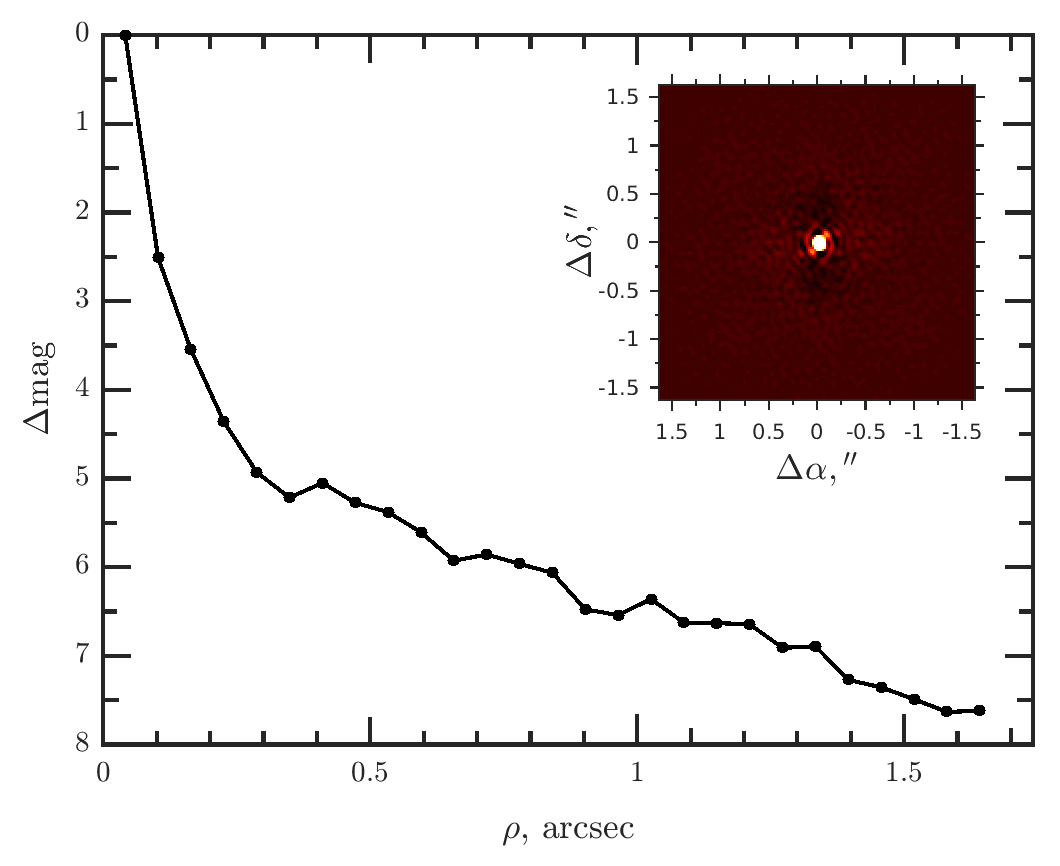}
    \caption{Contrast curve in $V$ band for TOI-6651 obtained from the speckle imager for PRL 2.5m telescope (on the left). The contrast curve from SPP speckle analysis in the $I_c$ band (on the right). The speckle ACF is displayed as an inset. No stellar companions are detected.}
    \label{fig:speckle}
\end{figure*}

Speckle imaging of TOI-6651 was conducted using the speckle imager on the PRL 2.5m telescope at Mount Abu Observatory in January 2024. The speckle imager is attached to one of the side ports of the telescope. The imager is equipped with a TRIUS PRO-814 CCD detector (model ICX814AL)\footnote{\url{https://www.sxccd.com/product/trius-sx814//}} that has a physical dimension of 3388 $\times$ 2712 pixels, each pixel measuring 3.69~$\mu$m $\times$ 3.69~$\mu$m. It offers a field of view (FOV) of 2.15$\arcmin$ $\times$ 1.7$\arcmin$, with a pixel scale of 38 $mas \ pixel^{-1}$, and acquires images in the $Bessel$-$V$ filter. A sequence of $\sim$2500 speckle frames of TOI-6651, each of 10~ms exposure time, were acquired.  The sky conditions during the observation were favorable, with an average seeing close to 1$\arcsec$. A custom data reduction pipeline written in {\tt PYTHON} was employed to analyze the speckle frames, following a procedure akin to methodologies outlined in \cite{2020ziegler} and \cite{2018tokovinin}. A brief outline of the steps is discussed below.

The speckle dataset is first dark-subtracted using a master dark derived from a series of dark frames obtained on the same night. Then, each of these dark-subtracted speckle images is cropped to a 128 $\times$ 128 pixel region centered on the image centroid. An average background value determined from the area outside the cropped region is subtracted from each frame. Following this, a normalized mean Power Spectrum Density (PSD) is estimated from the squared modulus of the Fourier Transform computed for each frame. The photon noise is determined from the PSD by averaging its values for spatial frequencies \(f > f_c\). The cutoff frequency is $f_c \approx \frac{D}{\lambda}$, where $D$ is the telescope diameter, and $\lambda$ is the central wavelength of the filter. This photon noise value is then subtracted from the observed PSD.

The pipeline next repeats the above steps for a set of speckle frames acquired for a standard star observed close to the target. The PSD of the science target is then divided by the PSD of the standard star to deconvolve the speckle transfer function (STF).  However, due to the unavailability of speckle imaging data for a standard star on the night of TOI-6651 observation, we employed a synthetic model of the STF for the deconvolution process, as discussed in \cite{2010Atokovinin}. The filtered observed PSD is then fitted with a model to extract both astrometric and photometric parameters \citep{2018tokovinin}. 

An autocorrelation function (ACF), also referred to as an autocorellogram, is computed from the PSD. A 5$\sigma$ contrast curve is then generated using this ACF. Concentric annuli centered on the primary star are defined within the ACF, and the detection limit is set as 5$\sigma$ brighter than the local mean within each annulus. The estimated contrast for TOI-6651 is $\Delta V=2.76$ and $\Delta V=4.98$ at 0.25$\arcsec$ and 1$\arcsec$, respectively. No stellar companions were detected for TOI-6651. Figure \ref{fig:speckle} shows the 5$\sigma$ contrast curve along with the ACF for TOI-6651.

\subsubsection{Speckle interferometry with SPP}

TOI-6651 was observed with the SPeckle Polarimeter (SPP; \cite{Strakhov23}) in the $I_c$ band on September 29, 2023. SPP is a facility instrument of the 2.5m telescope at the Caucasian Observatory of Sternberg Astronomical Institute (SAI) of Lomonosov Moscow State University. A fast low-noise CMOS Hamamatsu ORCA-quest was used as a detector, with a pixel scale of 20.6  $mas \ pixel^{-1}$, an angular resolution of 89 mas, and a field of view of 5$\arcsec\times5\arcsec$ centered on the star. The power spectrum was estimated from 4000 frames with 30 ms exposure and the atmospheric dispersion compensator was employed. We do not detect stellar companions, with detection limits of $\Delta I=4.7$ and $\Delta I=6.4$ at stellocentric distances of 0.25$\arcsec$ and 1.0$\arcsec$, respectively. The final contrast curve for TOI-6651 is shown in Figure \ref{fig:speckle}.

\subsection{Precise radial velocities}{\label{sec:paras2_obs}}

\begin{figure}
    \centering
    \includegraphics[width=\columnwidth]{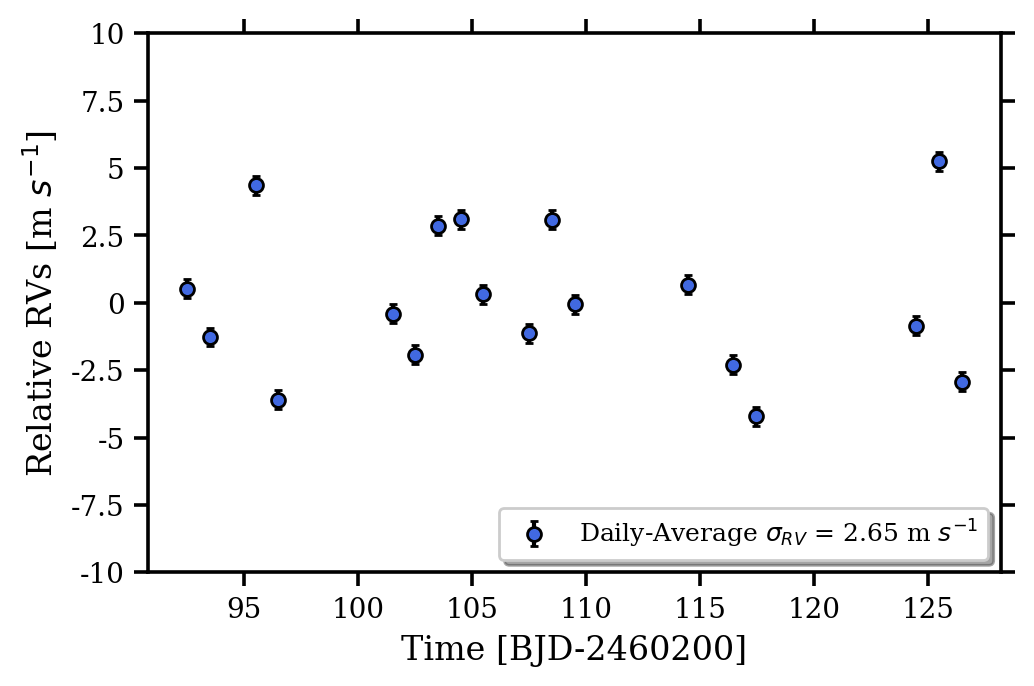}
    \caption{RVs of the standard star HD 109358 measured over a period of $\sim$35 days. The star has a V-band apparent magnitude of 4.25 and a spectral type of G0V.}
    \label{fig:rv_standard}
\end{figure}

The RV observations of TOI-6651 were made using PARAS-2 \citep{paras2_design,prl2.5m_and_paras2} spectrograph attached with the PRL 2.5m telescope at Mount Abu Observatory. PARAS-2 is a fiber-fed high-resolution (R$\approx$ 107,000) cross-dispersed echelle spectrograph maintained under highly stable conditions of temperature and pressure and works in the wavelength range of 3800-6900 $\AA$. The spectrograph uses the simultaneous referencing method using the Uranium Argon (UAr) hollow cathode lamp for wavelength calibration and precise RV calculations. Consistently, it has shown off-sky instrumental precision of $30-50 \ cm \ s^{-1}$ during the course of one night ($\sim$12 hours). One can refer to \citet{paras2_design,prl2.5m_and_paras2} for extensive details about the spectrograph. The methodology for the data reduction and RV calculations for the PARAS-2 spectra are described in Section~\ref{sec:paras2_pipeline}.
We took a total of 27 spectroscopic observations of TOI-6651 spanning 64 days from November 17, 2023, to January 20, 2024, all with an exposure time of 3600 seconds, with S/N at $5500~\AA$ ranging from 15 to 41 and mean S/N of 26. The corresponding photon noise in the RVs is in the range from 2.62 $m \ s^{-1}$ to 13.50 $m \ s^{-1}$ with a mean of 6.05 $m \ s^{-1}$, calculated using the techniques mentioned in \cite{Chaturvedi_2016, Chaturvedi_2018}. All the details of RVs and their corresponding errors without RV jitter are given in Table \ref{tab:rv_table}. Along with TOI-6651, we have also observed an RV standard star HD 109358 \citep{Bouchy2013} for a span of $\sim$35 days from December 13, 2023, to January 16, 2024, acquiring a total of 37 RVs. The measured daily RV dispersion is 2.65 $m \ s^{-1}$ (see Figure~\ref{fig:rv_standard}). 

\subsection{PARAS-2 data reduction and analysis pipeline}{\label{sec:paras2_pipeline}}

The PARAS-2 data reduction pipeline is written in IDL and is based on the PARAS-1 pipeline \citep{Chakraborty2014}, which uses the REDUCE package \citep{Piskunov2002} for extraction of cross-dispersed echelle spectra. The package has been adapted and evolved for the PARAS-2 in the same manner it is being done for the PARAS-1 pipeline.

Using the various calibration and science frames, the pipeline is capable of optimally extracting the spectra for science targets. In brief, a master bias frame is generated by a median combining all the bias frames, which is then subtracted from all non-biased frames. A master flat for each fiber is similarly created by median combining all the corresponding flat frames. These master images are further used for echelle order tracing, by finding the order curvature and its locations. The order tracing algorithm associated with REDUCE mainly uses a two-dimensional clustering process, in which it selects the probable order pixels and probes the level of clustering in each of them before merging the most probable orders selected. Since the instrument is very stable, we need not repeat this exercise on a regular basis, instead we use a master order trace. In case of large shifts in the cross-dispersion direction in the flat frames, we need to create the master trace again using the clustering method from REDUCE.

Before doing the optimal reduction, the pipeline does the cosmic ray rejection of the science frames using the Laplacian Edge detection method as described in \cite{Dokkum2001}. It is also important to estimate the scattered light beneath the orders extracted by the pipeline. For this purpose, the pipeline interpolates the background between orders after isolating the inter-order noise. A bleeding map of bright Argon lines is also created by acquiring the single fiber illumination with the UAr lamp. Finally, the spectra are extracted using the swath-based decomposition method.

The spectra need to be wavelength-calibrated for any further analysis. A coarse wavelength mapping of the spectrograph is done using the uranium linelist given in \cite{Ulines2021} by identifying their pixel positions on the detector. Then, for each order, a fourth-order polynomial is fitted between the central wavelength and the pixel position. This gives the wavelength solution for every order \citep{Ulines2021}, which is the blueprint of the wavelength solution, available at every instant. The relative wavelength calibration is done using this wavelength solution as the spectrograph is very stable. The wavelength-calibrated spectra of TOI-6651 are cross-correlated with a numerically weighted mask of G2 type star, similarly as described in \cite{Baranne1996} and \cite{Pepe_ccf_2002}.\\


\section{Result and analysis}\label{sec:analysis}

\subsection{Stellar parameters from spectral synthesis}\label{sec:paras_spec}

We estimated the stellar parameters of TOI-6651 using a combined spectrum with S/N $\approx 80$, which was generated by co-adding 8 individual high S/N spectra acquired with PARAS-2 (see Section~\ref{sec:paras2_obs}). The Zonal Atmospheric Stellar Parameters Estimator (\texttt{ZASPE}; \citealt{zaspe}) was used for spectral synthesis. \texttt{ZASPE} compares the co-added spectra with a library of synthetic spectra generated from ATLAS9 model atmospheres \citep{Castelli2003}, finding the best match through least square minimization. Additionally, a high S/N spectrum with S/N $\approx 50$ was obtained with the TRES instrument ($R \approx 44,000$) on the FLWO 1.5m telescope in September 2023, covering wavelengths from 3850-9096 $\AA$. The stellar parameters were derived using the Stellar Parameter Classification tool (SPC; \citealt{buchhave2010, buchhave2012, buchhave2014}) through cross-correlation with a grid of synthetic spectra based on Kurucz atmospheric models \citep{kurucz1992}. Both spectra were analyzed individually, and the weighted average of stellar parameter resulted in effective temperature ($T_{\rm eff}$) of $5933 \pm 57$ K, surface gravity ($\log{g}$) of $4.04 \pm 0.32$, metallicity ($[{\rm Fe/H}]$) of $0.213 \pm 0.030$ dex, and rotational velocity ($v\sin{i}_*$) of $4.88 \pm 0.14$ km s$^{-1}$. These parameters indicate that the host star is a sub-giant, metal-rich G-type star. We further use these parameters, along with the broadband spectral energy distribution (SED) and the evolutionary tracks, for a complete stellar characterization, which is elaborated in Sec~\ref{sec:global_modeling}.

\subsection{ Galactic kinematics}

We calculate the Galactic space velocity components $UVW$ in the barycentric frame utilizing the \texttt{gal$\_$uvw}\footnote{\url{https://pyastronomy.readthedocs.io/en/latest/pyaslDoc/aslDoc/gal_uvw.html}}function. These values are listed in Table~\ref{tab:star_table}, with UVW being positive in the directions of the Galactic center, Galactic rotation, and the north Galactic pole, respectively. We also provide these velocities with respect to the local standard of rest (LSR) using the solar velocities from \cite{Sch2010} (see Table~\ref{tab:star_table}). Based on our analysis, we found that TOI-6651 is associated with the thin disk \citep{Leggett1992, Bensby2014}. The BANYAN $\Sigma$ algorithm estimates cluster membership probabilities based on sky positions, proper motions, parallaxes, and RV, classifying TOI-6651 as field stars (>99$\%$ chance of membership) without associations to known young clusters \citep{Gagne2018}.

\subsection{Rotational period of the star}

Using the $v\sin{i}_*=4.88\pm0.14$ $km\ s^{-1}$, estimated with the spectral synthesis (see Section~\ref{sec:paras_spec}) and the stellar radius $R_*=1.721\pm0.065\ \rst$, the rotational period of the star is  $P_{rot}=17.9\pm0.8\ days$, calculated by assuming that star is observed equator-on. We used the archival long-term photometry data from Wide Angle Search for Planets (SuperWASP; \cite{Butters2010}) to find the periodically modulated signals attributed to the rotation of the host star. SuperWASP observations were conducted using a broadband filter with a passband ranging from 4000-7000 $\AA$. The dataset utilized for TOI-6651 in this study encompasses 13560 measurements spanning a four-year baseline from June 11, 2004, to July 23, 2008. Since the high cadence TESS data is available for only one sector, we did not account for it for Lomb-Scargle (LS). Due to significant noise in the SuperWASP data, the resulting periodogram is highly uncertain and difficult to interpret. We identified multiple peaks in the LS periodogram exceeding the 0.1\% FAP threshold. The most prominent peaks are clustered around a period of 16.5 days, consistent with the rotational period derived from the star's rotational velocity and radius.

\subsection{Periodogram analysis}

\begin{figure}[t!]
    \centering
    \includegraphics[width=\columnwidth]{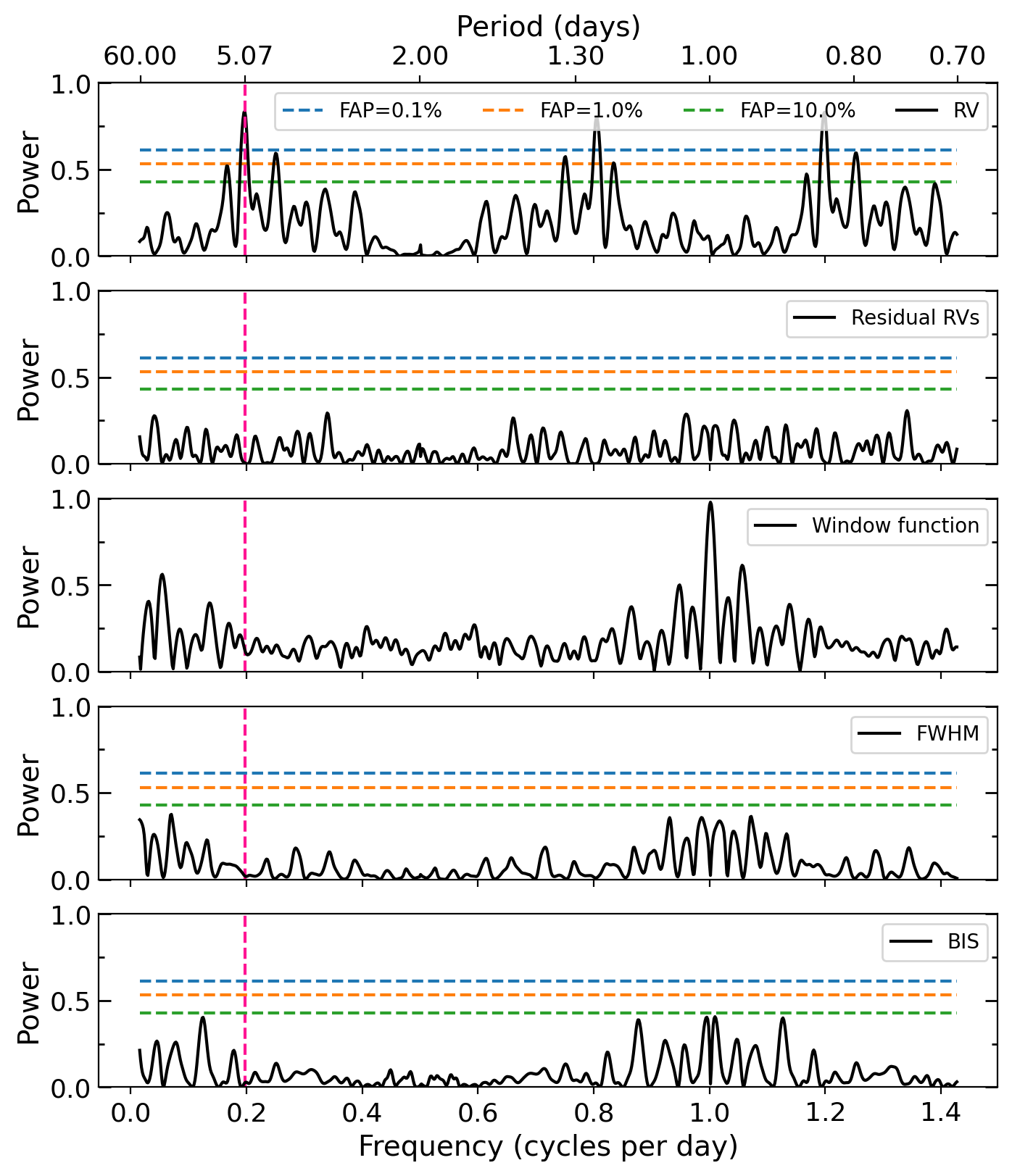}
    \caption{GLS periodogram for the RVs, residual RVs, window function, FWHM, and bisector span of TOI-6651 are shown in panels 1–5 (top to bottom), respectively. The most significant period in RVs is found to be 5.070 days, denoted by a pink dashed vertical line, which is consistent with the period 5.057 days found in the TESS transit signal. The FAP levels corresponding to 0.1\%, 1.0\%, and 10.0\% are shown in horizontal blue, orange, and green dashed lines, respectively.}
    \label{fig:periodogram}
\end{figure}

We compute the Generalized Lomb-Scargle (GLS) periodogram~\citep{periodogram} on PARAS-2 RV data (panel~\texttt{1} of Figure~\ref{fig:periodogram}) and calculate the theoretical   0.1$\%$,  1.0$\%$, and 10$\%$ false alarm probability (FAP) levels, indicated by the horizontal dotted lines. The baseline of our observations is $\sim$ 64 days, corresponding to a frequency resolution of 1/64 $\approx$ 0.015 d$^{-1}$. The most significant period is found to be 5.070 days, which is consistent with the period ($\approx$ 5.057 days) found in the TESS transit signal. The other two less significant peaks can be seen around $\approx$ 0.83 and $\approx$ 1.24 days, above the 0.1\% FAP level. After fitting for the 5.070~days signal, we do not observe any additional significant peak in the periodogram of residuals (panel~\texttt{2} of Figure~\ref{fig:periodogram}), suggesting that these two peaks are aliases of the most significant periodic signal. Furthermore, we compute the periodograms for the Window function, CCF FWHM, and bisector inverse slope (BIS) of PARAS-2 data, and these are shown in panels \texttt{3}, \texttt{4}, and \texttt{5} of Figure \ref{fig:periodogram}. These diagnostics serve as indicators of stellar activity, as they measure line asymmetries resembling Doppler shifts. However, our analysis did not reveal any statistically significant signals of stellar activity in the periodograms.

\begin{table}[t!]
\centering
\caption{Astrometry, photometry, and kinematics properties of TOI-6651.}
\begin{tabular}{lll}
\hline
\noalign{\smallskip}
 Parameter&Value& Ref.\\
\noalign{\smallskip}
\hline
\noalign{\smallskip}
Identifiers:\\
TIC&174328045&(1)\\
TYC&2286-515-1& (2)\\
2MASS&J01030925+3523208&(3)\\
\textit{Gaia}DR3&363314964357100800& (4)\\
\noalign{\smallskip}
Astrometry:\\
$\alpha_{J2000}$ & 01:03:09.248 & (4)\\
$\delta_{J2000}$ & +35:23:20.88 & (4)\\
$\mu_{\alpha}$ (mas yr$^{-1}$) & 10.544 $\pm$ 0.022 & (4)\\
$\mu_{\delta}$ (mas yr$^{-1}$) & -21.279 $\pm$ 0.017 & (4)\\
$\varpi^*$ (mas) &$4.727\pm0.021$ &(4)\\
$d$ (pc) &$211.53^{+0.94}_{-0.93}$ &(4)\\
\noalign{\smallskip}
Photometry$^*$:\\
$B_{T}$ & 11.024  $\pm$ 0.046 & (2)\\
$V_{T}$ & 10.315  $\pm$ 0.034 & (2)\\
$T$   & 9.6089 $\pm$ 0.0062 & (1)\\
$G$   & 10.045  $\pm$ 0.020 & (4)\\
$G_{BP}$   & 10.370 $\pm$ 0.020 & (4)\\
$G_{RP}$   & 9.558 $\pm$ 0.020 & (4)\\
$J$   & 9.022 $\pm$  0.029 & (3)\\
$H$   & 8.736 $\pm$  0.031 & (3)\\
$K_{S}$ & 8.684 $\pm$ 0.020 & (3)\\
$W1$  &  8.648 $\pm$ 0.030 & (5)\\
$W2$  &  8.691 $\pm$ 0.030 & (5)\\
$W3$  &  8.650 $\pm$ 0.030 & (5)\\
$W4$  &  8.658 $\pm$ 0.288 & (5)\\
\noalign{\smallskip}
Kinematics:\\
$U,V,W$ ($km \ s^{-1}$) &  3.783, -25.022, -11.639 & (6)\\
$[U,V,W]_{LSR}$ ($km \ s^{-1}$) &  14.883, -12.782, -4.389 & (6)\\
\noalign{\smallskip}
\hline
\noalign{\smallskip}
\multicolumn{3}{l}{\footnotesize{$^*$A systematic error floor has been applied to the uncertainties.}}\\
\multicolumn{3}{l}{\footnotesize{References: (1) \cite{2018AJ....156..102S}, (2) \cite{tycho},}}\\
\multicolumn{3}{l}{\footnotesize{ (3) \cite{JHK}, (4) \cite{gaiaedr3},}}\\
\multicolumn{3}{l}{\footnotesize{ (5) \cite{ALLWISE}, (6) This work}}\\
\end{tabular}
\label{tab:star_table}
\end{table}

\subsection{Global modeling}\label{sec:global_modeling}

\begin{figure}[t!]
    \centering
    \includegraphics[width=\columnwidth]{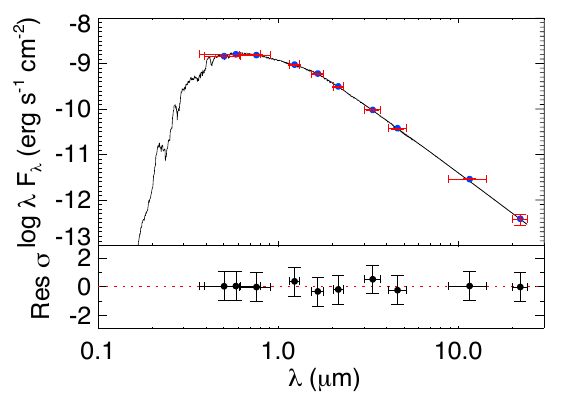}
    \caption{Spectral energy distribution of TOI-6651. The red symbols show the photometric measurements observed, while the horizontal bars indicate the effective width of the pass band. The blue points represent the model fluxes. The residuals are shown in the lower panel.}
    \label{fig:sed}
\end{figure}

We determined the stellar and planetary parameters of the TOI-6651 system with simultaneous fitting of the TESS LC and the PARAS-2 RV data using EXOFASTv2 \citep{exofast}, a global modeling software tailored for exoplanet fitting and developed in IDL. This software uses differential Evolution Markov Chain Monte Carlo (dMCMC) analysis to fit multiple datasets simultaneously. The fitting process continues until it meets the standard convergence criteria of EXOFASTv2: number of independent draws, $T_z$ > 1000 and Gelman–Rubin statistic $(GR) < 1.01$~\citep{Gelman_rubin1992,Gelman_rubin2006}.

\subsubsection{Stellar parameters}

To model the host star within the EXOFASTv2 framework, a combination of SED \citep{SED1} fitting using the Kurucz stellar atmosphere model \citep{Kurucz} and MESA Isochrones and Stellar Tracks (MIST) stellar evolutionary models \citep{mist_choi, mist_dotter} were employed. This combination, along with transit photometry, allows for precise estimation of the star's mass, radius, logg, and age \citep{Torres2008, Eastman2023}. Therefore, we used the TESS LC of TOI-6651 to model the host star. The broadband photometric magnitudes $G$, $G_{BP}$, $G_{RP}$ from \textit{Gaia}, $J, H, K$ from 2MASS, and $W1, W2, W3, W4$ from ALLWISE were incorporated into the SED fitting process, as given in Table~\ref{tab:star_table}. The uncertainties for these magnitudes are inflated to compensate for systematic floors in measuring accurate absolute photometry, as described in \cite{Lindegren_2021}. Priors were carefully selected to constrain the parameters effectively. Gaussian priors were set on $[{\rm Fe/H}]$ based on results from spectral synthesis (see Section~\ref{sec:paras_spec}) and on parallax measurements from \textit{Gaia} DR3 \citep{gaia2023} with systematic correction from \cite{Lindegren_2021}. We did not impose a prior constraint on $\log{g}$ and $T_{\rm eff}$ but provided the initial values derived from the spectral analysis (see Section~\ref{sec:paras_spec}). Additionally, a uniform prior was imposed to enforce an upper limit on V-band extinction from \citet{extinction} dust maps at the location of TOI-6651.

\begin{figure}[t!]
    \centering
    \includegraphics[width=0.99\columnwidth]{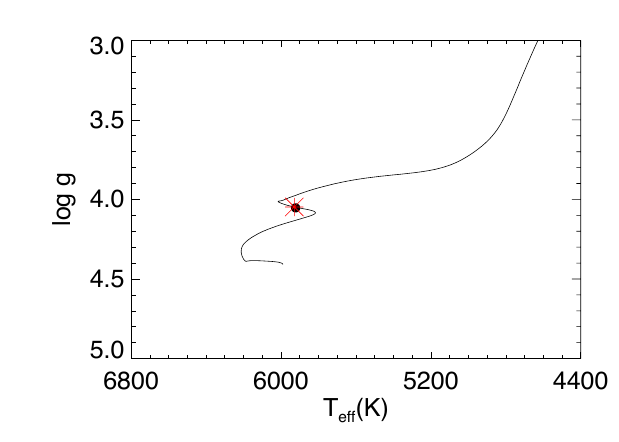}
    \caption{MIST evolutionary track for TOI-6651 shown as a solid black line. The black point indicates the $T_\mathrm{eff}$ and $\log{g}$, while the red asterisk denotes the current age of TOI-6651.}
    \label{fig:mist}
\end{figure}

The resulting posterior distribution of stellar parameters like $M_*$ and age show bimodality after all the chains converged. This bimodality arises due to the degeneracy between the MIST isochrones in the $T_{\rm eff}$ – $\log{g}$ plane (e.g., \cite{Carmichael_2021,Rodriguez_2021,toi1789}). The bimodal peaks for TOI-6651 occur at stellar masses of $1.323 \ M_\odot$ and $1.168 \ M_\odot$. We use the \texttt{splitpdf} tool within EXOFASTv2 to separate the posterior distribution into two solutions (referred to as high-mass and low-mass solutions) and report the corresponding parameters in Table~\ref{tab:planet_table}. The relative probabilities for the high-mass and low-mass solutions are 58.67\% and 41.33\%, respectively. We adopt the high-mass solution for our further analysis and discussion. The $T_\mathrm{eff}$ versus $\log{g}$ corresponding to the high-mass solution is marked in Figure~\ref{fig:mist}, where the most likely MIST evolutionary track for TOI-6651 is illustrated as a solid line. The resulting best-fit SED model with broadband photometry fluxes is shown in Figure~\ref{fig:sed}. The stellar parameters derived using EXOFASTv2, with median values and a 68\% confidence interval for each posterior (1-$\sigma$ uncertainty), are summarized in Table~\ref{tab:star_table}. From our analysis, we find the most probable host star parameters to be: $M_* = 1.323^{+0.050}_{-0.045} \ M_\odot$, $R_* = 1.721^{+0.065}_{-0.062} \ R_\odot$, $T_\mathrm{eff}=5940\pm110$ K, $\log{g} = 4.087^{+0.035}_{-0.032}~dex$, $[{\rm Fe/H}]=0.225^{+0.044}_{-0.045}$ at an age of $3.71^{+0.68}_{-0.78}$ Gyr. These parameters are consistent with those derived from spectral synthesis (see Section~\ref{sec:paras_spec}).

\subsubsection{Planetary parameters}

\begin{figure*}[t!]
    \centering
    \includegraphics[width=0.99\columnwidth]{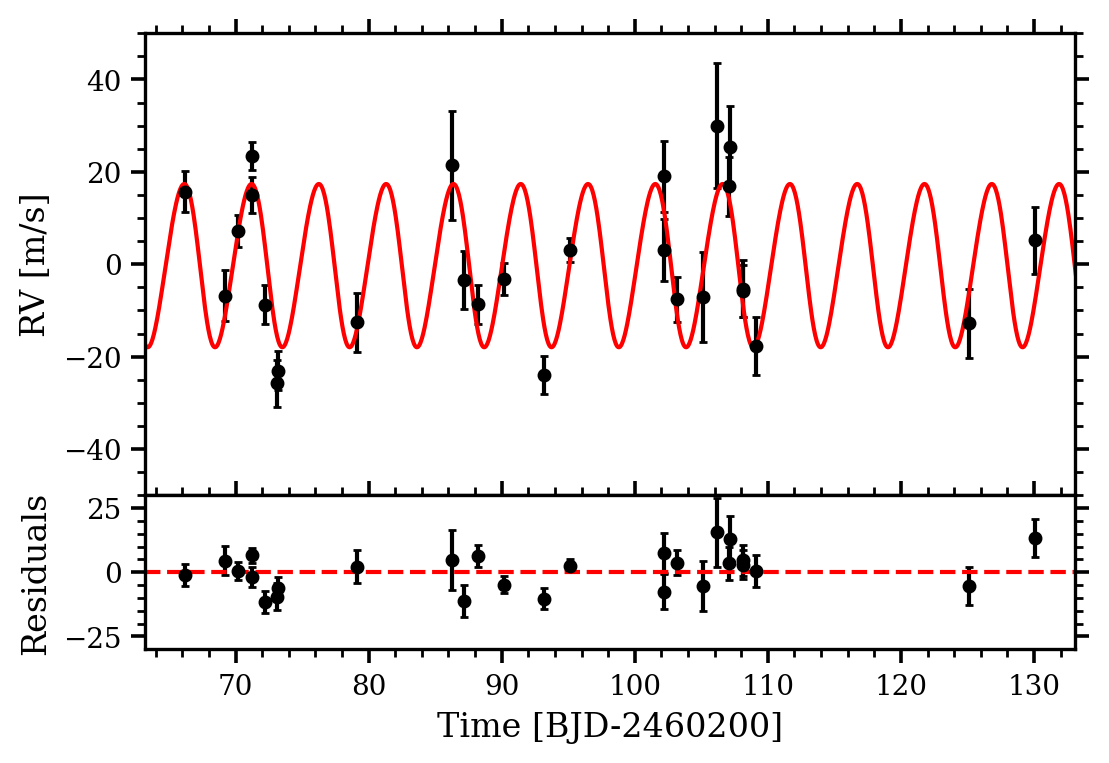}
    \hspace{0.3cm}
    \includegraphics[width=0.99\columnwidth]{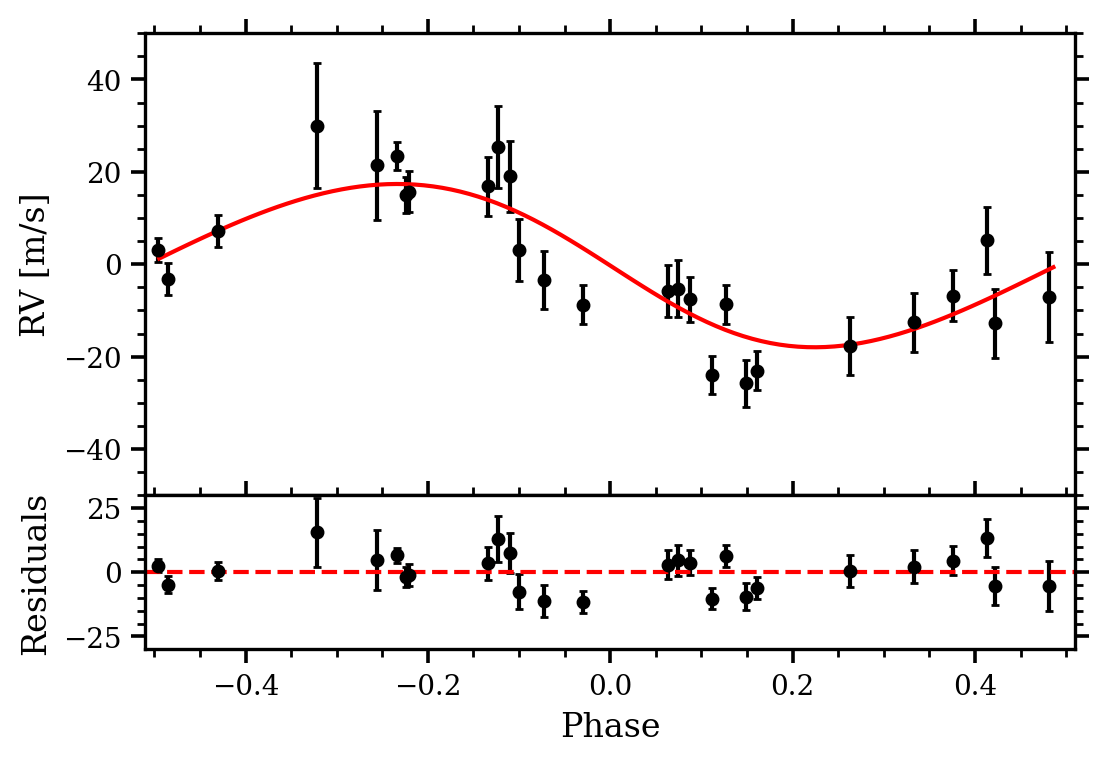}
    \caption{TOI-6651 RVs (black dots) from 27 spectroscopic observations with PARAS-2 are plotted with time (on the left). TOI-6651 RVs plotted with orbital phase (on the right). In both plots, the best-fit RV model with EXOFASTv2 is represented by the red line, and residuals between the best-fit model and the data are shown in their respective lower panel.}
    \label{fig:rv}
\end{figure*}

The joint modeling of TESS photometry with PARAS-2 RVs is done by keeping all orbital fitting parameters (such as $b$, $i$, $a$, $R_{p}$, $K$, $e$, and $\omega$) as free variables and only supplying initial values for $P$ and $T_{c}$, derived from the TESS SPOC pipeline. The transit model is generated using the method outlined by \citet{Mandel2002}, while the RV data are modeled with a standard non-circular Keplerian orbit. We use the default quadratic limb-darkening law for the TESS passband, and its coefficients ($u_{1}$ and $u_{2}$) are computed based on tables reported in \citet{Claret} and \citet{Claret_tess}. TESS observations of TOI-6651 are from two sectors with different exposure times, so we resampled the transit model for sector 17 LC, which is 1800-second exposures over 10 steps to account for smearing. The joint modeling reveals a sub-Saturn planet with a most probable radius of $R_P = 5.09^{+0.27}_{-0.26} \ R_\oplus$ and a mass of $M_P = 61.0^{+7.6}_{-7.9} \ M_\oplus$, corresponding to the RV semi-amplitude of $K = 18.9 \pm 2.3 \ \mathrm{m \ s^{-1}}$ in an eccentric orbit ($e = 0.091^{+0.096}_{-0.062}$). The best-fitted orbital and planetary parameters for both high and low mass solutions are listed in Table~\ref{tab:planet_table} with their median values and 68\% confidence interval of the posterior (1-$\sigma$ uncertainty). The TESS LCs and PARAS-2 RVs with their best-fitted models are shown in Figures~\ref{fig:tesslc} and \ref{fig:rv}, respectively.

\renewcommand{\arraystretch}{1.12}
\begin{table*}[b!]
\caption{Summary of EXOFASTv2 fitted and derived parameters with 68\% confidence interval for TOI-6651 system.}
\label{tab:planet_table}   
\centering
\begin{tabular}{lllll}
\hline
\noalign{\smallskip}
Parameter&Description&Value ($\mathds{P}=58.67\%)^{1}$ & Value ($\mathds{P}=41.33\%)^{1}$\\
\noalign{\smallskip}
\hline
\noalign{\smallskip}
\multicolumn{2}{l}{Stellar Parameters:}\\
$M_*$ &Mass (\msun) &$1.323^{+0.050}_{-0.045}$ &$1.168^{+0.036}_{-0.042}$\\
$R_*$ &Radius (\rsun) &$1.721^{+0.065}_{-0.062}$ &$1.772^{+0.067}_{-0.066}$\\
$L_*$ &Luminosity (\lsun) &$3.33^{+0.11}_{-0.12}$ &$3.30^{+0.12}_{-0.13}$\\
$\rho_*$ &Density (cgs) &$0.366^{+0.045}_{-0.037}$ &$0.296^{+0.035}_{-0.032}$\\
$\log{g}$ &Surface gravity (cgs) &$4.087^{+0.035}_{-0.032}$ &$4.008^{+0.033}_{-0.035}$\\
$T_{\rm eff}$ &Effective temperature (K) &$5940\pm110$ &$5840\pm110$\\
$[{\rm Fe/H}]$ &Metallicity (dex) &$0.225^{+0.044}_{-0.045}$ &$0.212^{+0.045}_{-0.046}$\\
$Age$ &Age (Gyr) &$3.71^{+0.68}_{-0.78}$ &$6.60^{+1.1}_{-0.79}$\\
$EEP$ &Equal Evolutionary Phase &$401^{+14}_{-21}$ &$452.2^{+5.4}_{-5.1}$\\
$A_V$ &V-band extinction (mag) &$0.109^{+0.024}_{-0.043}$ &$0.096^{+0.032}_{-0.049}$\\
$\sigma_{SED}$ &SED photometry error scaling  &$0.335^{+0.13}_{-0.080}$ &$0.345^{+0.13}_{-0.083}$\\
$v\sin{i_*}$ & Projected rotational velocity$^{2}$ ($km \ s^{-1}$) & {$4.88\pm0.14$}& {$4.88\pm0.14$}\\
\noalign{\smallskip}
\multicolumn{2}{l}{Planetary Parameters:}\\
$P$ &Period (days) &$5.056973^{+0.000016}_{-0.000018}$ &$5.056973^{+0.000016}_{-0.000018}$\\
$R_P$ &Radius $(R_\oplus)$ &$5.09^{+0.27}_{-0.26}$ &$5.31\pm 0.28$\\
$M_P$ &Mass $(M_\oplus)$ &$61.0^{+7.6}_{-7.9}$ &$56.3^{+7.0}_{-7.3}$\\
$T_C$ &Time of conjunction (\bjdtdb) &$2459857.7116\pm0.0014$ &$2459857.7116^{+0.0014}_{-0.0015}$\\
$a$ &Semi-major axis (AU) &$0.06330^{+0.00078}_{-0.00072}$ &$0.06073^{+0.00062}_{-0.00074}$\\
$i$ &Inclination (Degrees) &$84.93^{+0.63}_{-0.46}$ &$84.09^{+0.50}_{-0.48}$\\
$e$ &Eccentricity  &$0.091^{+0.096}_{-0.062}$ &$0.089^{+0.094}_{-0.060}$\\
$\omega_*$ &Arg of periastron (Degrees) &$116^{+62}_{-38}$ &$117^{+62}_{-39}$\\
$T_{\rm eq}$ &Equilibrium temp$^{3}$ (K) &$1493^{+14}_{-15}$ &$1522\pm15$\\
$K$ &RV semi-amplitude (m/s) &$18.9\pm2.3$ &$18.9^{+2.2}_{-2.3}$\\
$R_P/R_*$ &Radius of planet in stellar radii  &$0.02715^{+0.00075}_{-0.00078}$ &$0.02753^{+0.00076}_{-0.00079}$\\
$a/R_*$ &Semi-major axis in stellar radii  &$7.91^{+0.31}_{-0.28}$ &$7.37^{+0.28}_{-0.27}$\\
$\delta$ &$\left(R_P/R_*\right)^2$  &$0.000737^{+0.000041}_{-0.000042}$ &$0.000758\pm0.000043$\\
$\delta_{\rm TESS}$ &Transit depth in TESS (frac) &$0.000784\pm0.000034$ &$0.000787^{+0.000034}_{-0.000033}$\\
$\tau$ &In/egress transit duration (days) &$0.0070^{+0.0014}_{-0.0016}$ &$0.0082^{+0.0016}_{-0.0018}$\\
$T_{14}$ &Total transit duration (days) &$0.1504^{+0.0032}_{-0.0029}$ &$0.1517^{+0.0033}_{-0.0031}$\\
$b$ &Transit impact parameter  &$0.662^{+0.062}_{-0.12}$ &$0.718^{+0.049}_{-0.091}$\\
$\rho_P$ &Density (cgs) &$2.52^{+0.52}_{-0.44}$ &$2.04^{+0.42}_{-0.36}$\\
$logg_P$ &Surface gravity (cgs) &$3.360^{+0.065}_{-0.069}$ &$3.287^{+0.065}_{-0.070}$\\
$\fave$ &Incident Flux (\fluxcgs) &$1.112^{+0.047}_{-0.051}$ &$1.201^{+0.051}_{-0.055}$\\
$T_P$ &Time of Periastron (\bjdtdb) &$2459852.93^{+0.82}_{-0.41}$ &$2459852.94^{+0.84}_{-0.42}$\\
$e\cos{\omega_*}$ &  &$-0.029^{+0.039}_{-0.055}$ &$-0.029^{+0.039}_{-0.054}$\\
$e\sin{\omega_*}$ &  &$0.060^{+0.11}_{-0.064}$ &$0.057^{+0.11}_{-0.062}$\\
$M_P\sin i$ &Minimum mass $(M_\oplus)$ &$60.7^{+7.6}_{-7.9}$ &$56.0^{+7.0}_{-7.3}$\\
$M_P/M_*$ &Mass ratio  &$0.000138^{+0.000017}_{-0.000018}$ &$0.000145\pm0.000018$\\
\noalign{\smallskip}
\multicolumn{2}{l}{Wavelength Parameters (TESS):}\\
$u_{1}$ &Linear limb-darkening coeff  &$0.281\pm0.038$ &$0.294\pm0.039$\\
$u_{2}$ &Quadratic limb-darkening coeff  &$0.288\pm0.036$ &$0.281\pm0.036$\\
\noalign{\smallskip}
\multicolumn{2}{l}{Telescope Parameters (PARAS-2):}\\
$\gamma_{\rm rel}$ &Relative RV Offset (m/s) &$-0.8\pm1.5$ &$-0.8\pm1.5$\\
$\sigma_J$ &RV Jitter (m/s) &$5.4^{+1.8}_{-1.5}$ &$5.4^{+1.8}_{-1.5}$\\
$\sigma_J^2$ &RV Jitter Variance  &$28^{+22}_{-14}$ &$28^{+22}_{-14}$\\
\noalign{\smallskip}
\multicolumn{2}{l}{Transit Parameters (TESS):}& sector 17, sector 57& sector 17, sector 57\\
$\sigma^{2}$ &Added Variance  &$(1.10^{+0.31}_{-0.29}, 1.89^{+0.54}_{-0.53})\times 10^{-8}$ & $(1.10^{+0.31}_{-0.29}, 1.89^{+0.54}_{-0.53})\times10^{-8}$\\
\noalign{\smallskip}
\hline
\noalign{\smallskip}
\multicolumn{4}{l}{\footnotesize{$^1$Relative probability, $^2$Spectroscopically derived (see Section~\ref{sec:paras_spec}),  $^{3}$Assumes no albedo and perfect redistribution}}\\
\end{tabular}
\end{table*}

\subsection{Tidal circularization}\label{tidal}

We estimated the timescales for tidal circularization using equation 2 from \cite{circ_time}, and considering the tidal quality factor $Q_p = 10^{5}$, which is within the range of values estimated for the Neptune to Saturn-like planets \citep{Lainey2009, subjak2022}. The stellar tidal quality factor influences the circularization timescale only if the value of $Q_p$ is larger than $10^{5.5}$ \citep{Jackson_2008}. For TOI-6651b, the estimated circularization timescale is around $4.4\pm1.6$ Gyr. Compared to the age of the host star $3.71^{+0.68}_{-0.78}$ Gyr, these estimates suggest that TOI-6651b could maintain its eccentric orbit to the present day.


\section{Discussion}\label{sec:discussion}

\subsection{Mass radius diagram and internal structure}

\begin{figure}[t!]
\centering
\includegraphics[width=\columnwidth]{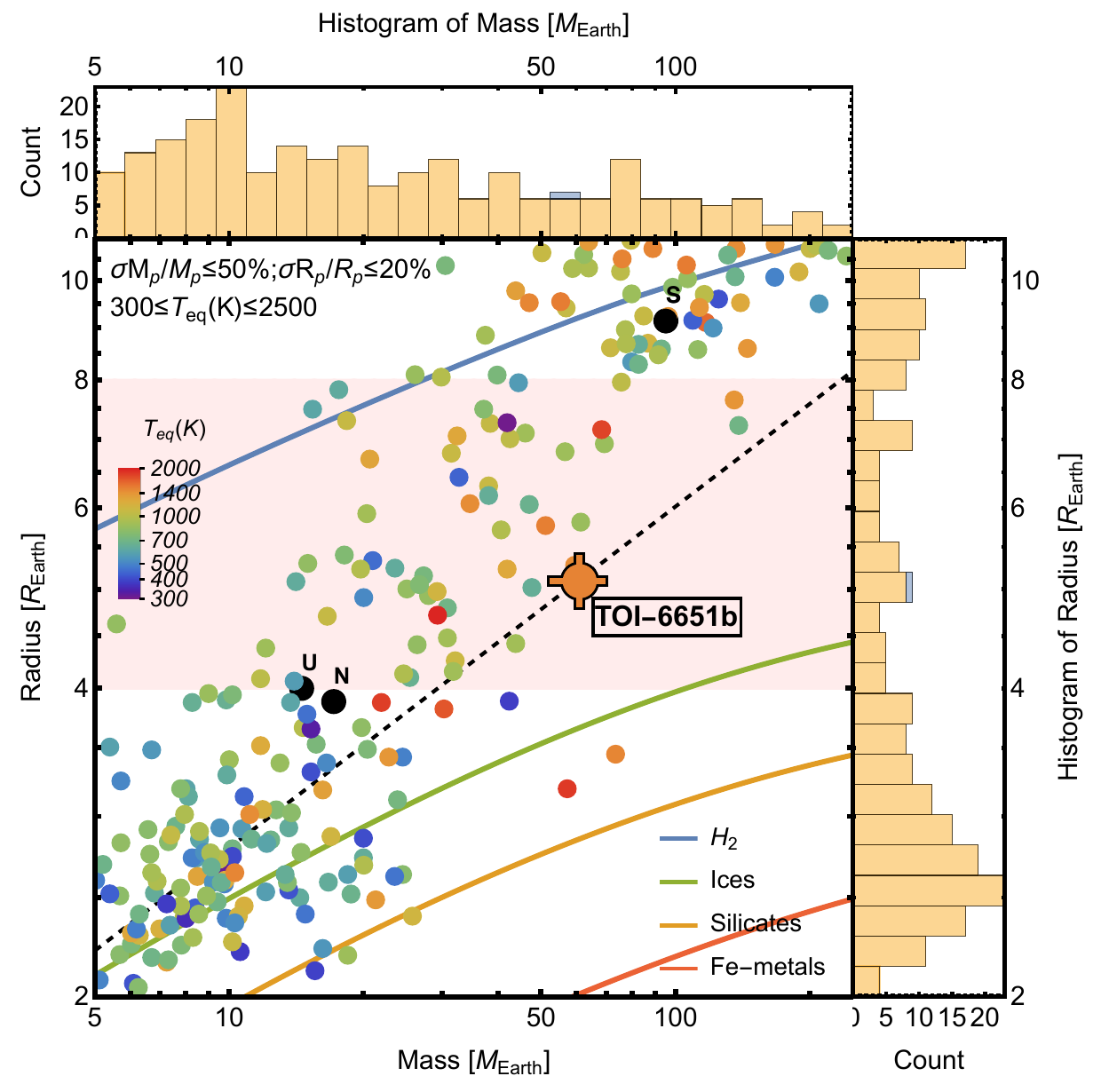}
    \caption{Planetary radius vs. mass of known exoplanets, with histograms in the right and top panels depicting the distribution of radii and masses, respectively. The sample is from the TepCat database \citep{TEPcat} and includes exoplanets with mass and radius constraints better than 50\% and 20\%, respectively, and equilibrium temperatures of planets ($T_{eq}$) between 300 K and 2500 K. The light pink highlighted region represents the sub-Saturn classification, while the black dashed line represents the iso-density curve for $\rho_P=2.52$. The red, orange, green, and blue solid curves represent the M-R diagram for pure-iron, pure-rock, pure high-pressure ices \citep{Zeng_2021} and pure $H_2$ core composition \citep{Becker_2014}, respectively. The Solar system planets are represented as black dots. TOI-6651b is positioned above the pure ice composition, indicating the presence of a significant fraction of a H/He envelope.}
    \label{fig:massradius}
\end{figure}

We plot the mass-radius (M-R) diagram for known exoplanets and display the position of TOI-6651b in Figure \ref{fig:massradius}. The sample of exoplanets is taken from the \texttt{TepCat} database \citep{TEPcat}, and we only select planets with equilibrium temperature ($T_{eq}$) between 300~K and 2500~K, and with mass and radius constraints better than 50\% and 20\%, respectively. We have also plotted the M-R curves for pure iron, pure rock, pure high-pressure ices \citep{Zeng_2021}, and pure $H_2$ core composition \citep{Becker_2014}.

It can be seen in the Figure that TOI-6651b occupies a position between pure high-pressure ices and pure $H_2$ core compositions, suggesting the presence of a significant envelope mass fraction ($f_{env}$). Additionally, its high density ($\rho_P = 2.52^{+0.52}_{-0.44}\ g\ cm^{-3}$) positions it among the densest known sub-Saturns, indicating a massive core and a smaller $f_{env}$. We used the \texttt{photoevolver}\footnote{\url{https://github.com/jorgefz/photoevolver}} python code \citep{Fernandez_23} to model the internal structure of TOI-6651b. To first order, sub-Saturns can be approximated as two-component planets consisting of a rocky heavy-element core surrounded by an H/He envelope, as the derived $f_{env}$ is not sensitive to the detailed composition of the planet core \citep{Petigura_2016}. The \texttt{photoevolver} code provides the models (functions) that define the planet's internal structure as well as its evolution. For internal structure, it defines four crucial parameters: the radius ($R_{core}$) and mass ($M_{core}$) of the core, and the radius ($R_{env}$) and $f_{env}$ of the surrounding envelope. These parameters are related to the total radius $R_P$ and mass $M_P$ of the planet as $R_P=R_{core}+R_{env}$ and $f_{env}=\frac{M_P-M_{core}}{M_P}$.

We utilized the empirical M-R relationships provided by \cite{Otegi20} to infer the core M-R relationship. For the H/He envelope, we adopted the envelope structure model of \cite{Chen_2016} to relate the $R_{\text{env}}$ and $f_{\text{env}}$ of the surrounding envelope. Given the total mass and radius of the planet and the irradiation flux, the \texttt{photoevolver} employs the above equations and model relations to solve for the internal structure parameters at the given system age. The resulting internal structure of TOI-6651b is summarized in Table \ref{tab:internal_structure}. These results indicate that $\approx$ 87\% of the planet's mass consists of dense materials such as rock and iron in the core, with a core radius of $3.26\pm0.19$ $R_\oplus$, which occupies only about 26\% of the planet's volume. The remaining mass comprises a low-density envelope of hydrogen and helium, which occupies most of the planet's volume.

\begin{table}[t!]
    \centering
    \caption{Internal structure of TOI-6651b.}
    \begin{tabular}{lll}
         \hline
         \noalign{\smallskip}
         Parameter & Description & Value\\
         \noalign{\smallskip}
         \hline
         \noalign{\smallskip}
         $M_{core}$ & Core mass $(M_\oplus)$ &  $53.3\pm7.3$\\
         $R_{core}$ & Core radius $(R_\oplus)$ & $3.26\pm0.19$\\
         $R_{env}$ & Envelope radius $(R_\oplus)$ & $1.83\pm0.33$\\
         $f_{env}$ & Envelope mass fraction & $0.126\pm0.037$\\
         \noalign{\smallskip}
         \hline
    \end{tabular}
    \label{tab:internal_structure}
\end{table}

\subsection{The formation scenario}\label{sec:formation}

TOI-6651b has a core mass of $\approx$ 53$M_\oplus$ (87\% of the total mass), making it massive enough to trigger runaway gas accretion and become a gas giant planet \citep{POLLACK_1996}. If this were the case, then TOI-6651b might have undergone significant atmospheric mass loss (see Section~\ref{sec:intro}) earlier in its history. However, due to its high core mass, it is expected to be immune to significant photoevaporation \citep{Petigura_2017}, wherein a star's radiation (X-ray and EUV) strips away a planet's atmosphere \citep{Owen_2017}. The other possibility is that the planet's orbit became highly elliptical, and it underwent high-eccentricity migration, during which the orbit was circularized by tidal damping at periastron, leading to the loss of its atmosphere through tidal heating (Tidal Stripping; \citet{Owen2018}; \citet{Matsakos_2016}; \citet{Vick_lai_2019}). Dense and massive sub-Saturns like TOI-6651b are more likely to lose their atmosphere due to tidal heating compared to photoevaporation, which is typically the dominant mechanism for less massive and less dense sub-Saturns \citep{Vissapragada2022}.

It has also been proposed that the formation of massive dense sub-Saturns is likely the result of dynamical instabilities \citep{Petigura_2017, Brady_2018}, and TOI-6651b may never have undergone runaway accretion of gas. Initially, TOI-6651b might have had neighboring planets but was situated in a dynamically unstable architecture, eventually resulting in close encounters and planet-planet scattering. Given its close proximity to the star ($P \approx 5$ days), TOI-6651b is deeply embedded in the gravitational field of its host star, making scattering events more likely to result in mergers rather than ejections from the system. The maximum velocity a planet can impart to its neighbor is the escape velocity \citep{Goldreich_04}. For TOI-6651b, we calculated the escape velocity and the orbital velocity to be $38.8\pm2.7$ km/s and $137\pm3$ km/s, respectively. Since the escape velocity is smaller than the orbital velocity (ratio $\approx$ 0.28), single scattering events cannot lead to ejections. Therefore, any previous dynamical instabilities would likely have caused planet mergers, increasing the total mass of TOI-6651b, and the slightly eccentric orbit today might be a relic of that process.

Another possible scenario involves the merger of several sub-critical cores during the disk dispersal stage, and the resulting core accreted the envelope from the gas-depleted disk. Even if the merger took place after the gas disk was fully dispersed and sub-critical cores had their own gas envelopes, then under certain conditions, mergers can form a massive core by breaking apart the lighter bound gaseous envelopes of sub-critical cores \citep{Liu_2015}. The formation of super-massive Neptune-sized planet TOI-1853b, through mergers and scattering, was also presented by \cite{Naponiello_2023}. Although it is difficult to ascertain the precise detailed formation history of the planet, further detection and in-depth studies of such planets can provide more insight into their origins.

\subsection{Position in the Neptunian desert}

\begin{figure}[t!]
    \centering
    \includegraphics[width=\columnwidth]{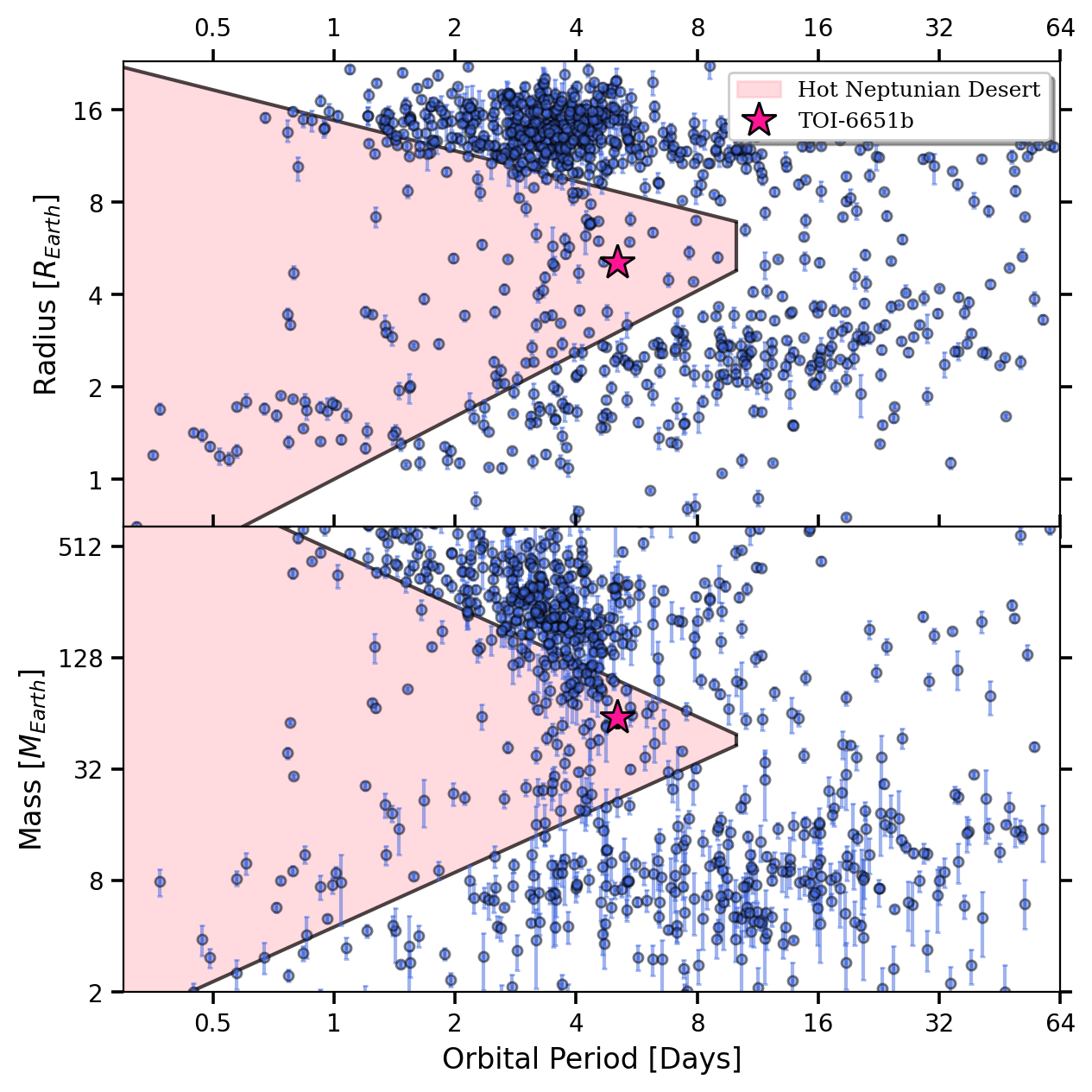}
    \caption{Planetary radii vs. orbital period (upper panel) and planetary masses vs. orbital period (lower panel) of known exoplanets from the NASA Exoplanet Archive \citep{NASA_EXO_Archive_Akeson_2013} with mass and radius constraints better than 50\% and 20\%, respectively. The light pink highlighted region represents the hot Neptunian desert with boundaries defined by \cite{Mazeh2016}. The position of TOI-6651b is shown as a pink star.}
    \label{fig:neptunedesert}
\end{figure}

\begin{figure*}[t!]
    \centering
    \includegraphics[width=0.97\columnwidth]{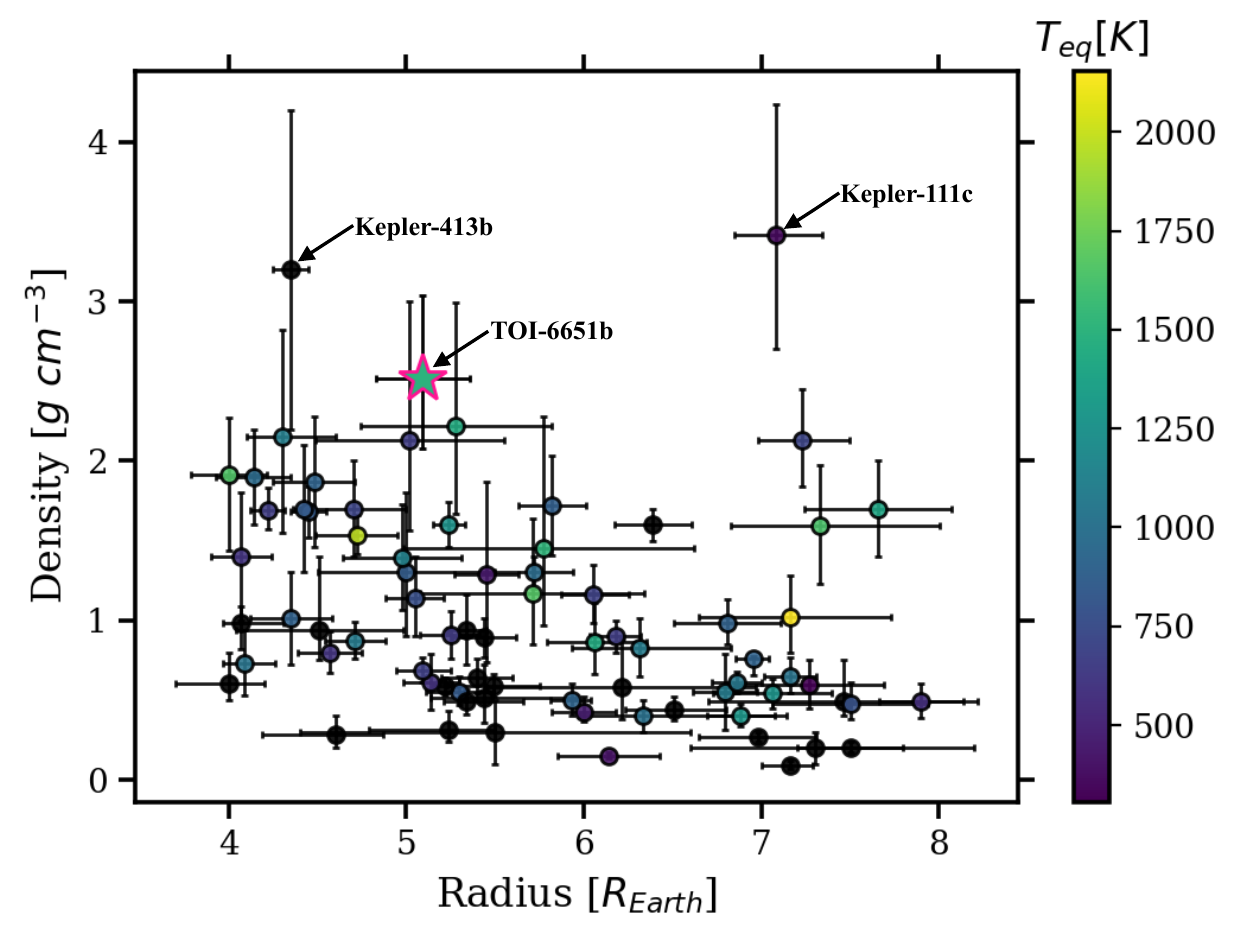}
    \hspace{0.3cm}
    \includegraphics[width=0.99\columnwidth]{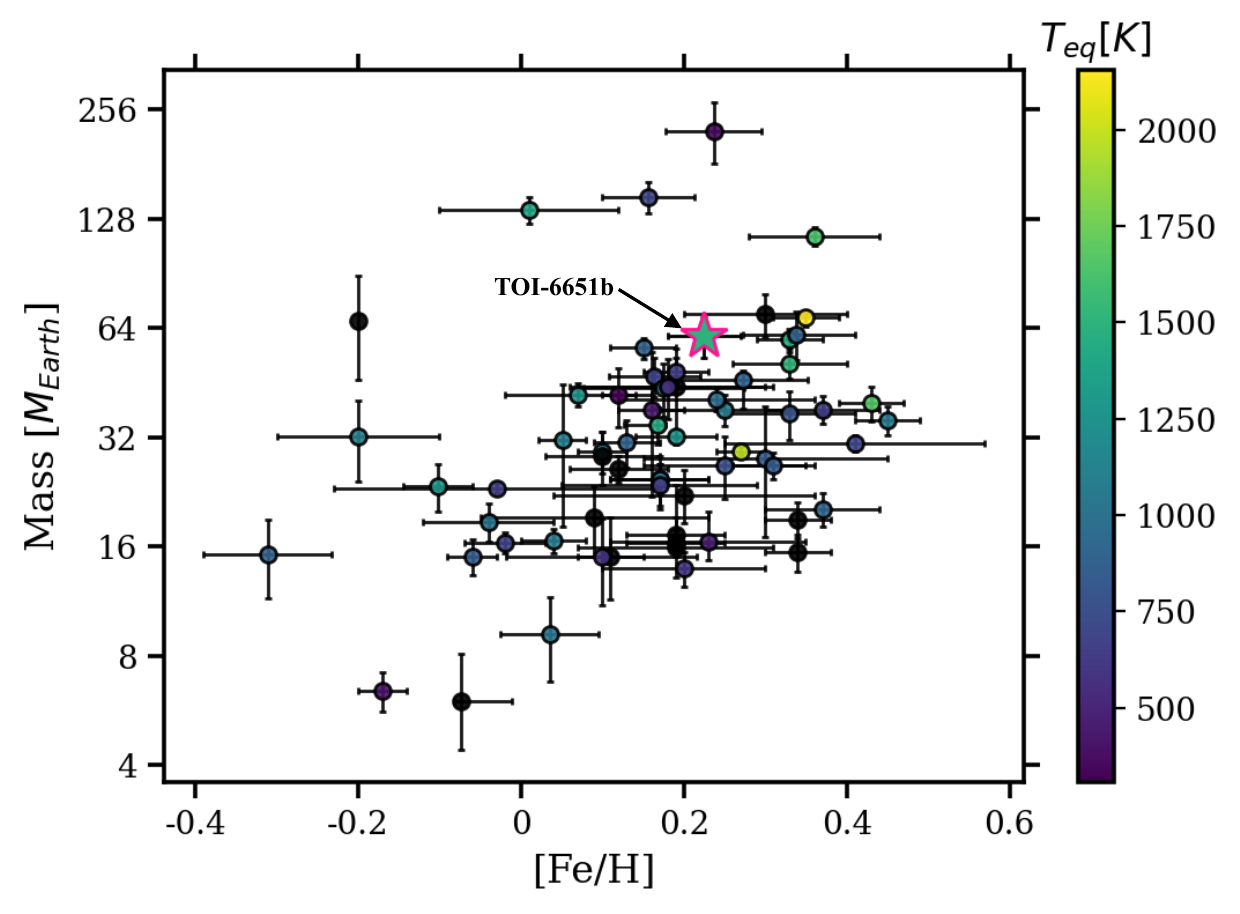}
    \caption{Bulk density of known sub-Saturns plotted against their radius (on the left), with TOI-6651b being the third-highest density among them (after Kepler-111c \cite{Dalba_2024}, and Kepler-413b \cite{Kostov_2014}). Mass of the sub-Saturns as a function of their host star metallicity (on the right), showing a positive correlation between them. The color of individual points is according to the equilibrium temperature of the planet. The position of TOI-6651b in both plots is shown as a pink edge-colored star.}
    \label{fig:sub_saturns_stat}
\end{figure*}

The hot Neptunian desert (also known as the sub-Jovian desert) is a region of radius-period and mass-period parameter space where sub-Jovian planets with $P<10$ days have been rarely found, and this scarcity cannot be attributed to observational biases \citep{szabo2011}. The probable mechanism for the origin of the Neptunian desert is atmospheric mass loss via photoevaporation and tidal heating following high-eccentricity migration. TOI-6651b ($P \approx 5.06$ days) resides at the edge of the desert, considering the boundary up to an orbital period of $\sim$ 5 days. The latter part of the desert, spanning orbital periods from 5 to 10 days, lacks a comprehensive explanation \citep{Mazeh2016,Hartman2019,Esposito2019,subjak2022,Osborn2023}, which makes TOI-6651b a compelling candidate for studying the origin of the Neptunian desert.

Figure~\ref{fig:neptunedesert} illustrates the position of TOI-6651b in both the mass-period and radius-period parameter spaces along with other known exoplanets, as well as the Neptunian desert with boundaries defined by \cite{Mazeh2016}. We find that TOI-6651b is close to the upper boundary of the desert in the mass-period parameter space. Various theories are proposed to elucidate the boundaries of the desert \citep[e.g.,][]{Owen2018,Vissapragada2022,Thorngren2023}. \cite{Owen2018} argue that photoevaporation may account for the lower boundary, while tidal effects following high-eccentricity migration could be responsible for the upper boundary. However, \cite{Thorngren2023} suggests a slightly different perspective, indicating that photoevaporation may partially contribute to carving out the upper boundary. While the detailed discussion of these theories falls beyond the scope of our current study, we contribute by presenting a new hot sub-Saturn located near the upper boundary of the mass-period desert, with a precise mass determination of $\approx$ 13$\%$. This finding will be useful for future studies on the functional dependence of the upper boundary on stellar and planetary properties.

\subsection{Position of TOI-6651b in sub-Saturns}

In this section, we compare TOI-6651b in the context of other known sub-Saturns. We used data from the NASA Exoplanet Archive \citep{NASA_EXO_Archive_Akeson_2013}, and if a planet is studied multiple times, we selected the parameters from the study that provided the most robust constraints on basic planetary parameters. So far\footnote{According to the NASA Exoplanet Archive \citep{NASA_EXO_Archive_Akeson_2013} as of July 17, 2024}, 74 sub-Saturns have been discovered, with radius and mass constraints better than 20 and 50 percent, respectively. Notably, 22 of these discoveries were made with TESS. Among these sub-Saturns, 41 are classified as ``hot sub-Saturns'' ($P \leq 10$ days). Furthermore, a select few sub-Saturns have densities exceeding $2.0\ g\ cm^{-3}$, a rarity in this class of exoplanets. Among these dense sub-Saturns, TOI-6651b emerges as the third-highest density sub-Saturn (after Kepler-111c \cite{Dalba_2024} and Kepler-413b \cite{Kostov_2014}) and the most dense among those detected with TESS. The density of known sub-Saturns as a function of their radius is shown in the left panel of Figure \ref{fig:sub_saturns_stat}.

Massive sub-Saturns tend to be found around metal-rich stars, as there were more solids present in the protoplanetary disk to form a massive core. Alternatively, slightly more massive cores can lead to dynamical instabilities, which may result in mergers that form even more massive cores capable of accreting larger gaseous envelopes \citep{Lee_2015,Petigura_2017}. The correlation between the mass of sub-Saturns and the metallicity of their host stars has been previously observed and discussed by \cite{Petigura_2017}. In the right panel of Figure \ref{fig:sub_saturns_stat}, we plot the mass of sub-Saturns as a function of their host star's metallicity. We found that the $M_P$ - $[Fe/H]$ positive correlation is consistent for the sub-Saturns detected so far, including the recent discoveries from TESS, in addition to those from Kepler and K2.

\subsection{Future follow-up opportunities}

Given the equilibrium temperature of planet $T_{eq}=1493\pm15\ K$ and the surface gravity $g_P=22.9\pm3.6\ m\ s^{-2}$, along with the assumption of a $H/He$ atmosphere with a mean molecular mass of $\mu=2.3\ amu$, the atmospheric scale height is calculated to be $H_b=235\pm37 km$ using the formula $H_b = \frac{k T_{eq}}{\mu g_P}$ \citep{Madhusudhan_2014}. TOI-6651b has a low Transmission Spectroscopic Metric (TSM, \cite{Kempton_2018}) of $19.6\pm4.3$. With a low transit depth of 0.737 ppt and a slowly rotating host with $v\sin{i}_*=4.88\ km\ s^{-1}$, the Rossiter–McLaughlin (RM) effect (\cite{Rossiter_1924}, \cite{McLaughlin_1924}) semi-amplitude for TOI-6651b is estimated to be $\approx$ $2\ m\ s^{-1}$. Further RV observations and transit-timing variations (TTV) from upcoming TESS data can be used to check for the presence of other planets in the system.


\section{Summary}\label{sec:summary} 

We presented the discovery and characterization of a dense sub-Saturn exoplanet, TOI-6651b, transiting a metal-rich G-type sub-giant star. Our analysis combined RV data from the PARAS-2, transit photometry data from TESS, and speckle observations from the PRL and SAI telescopes. This enabled us to establish the planetary nature of TOI-6651b and determine its mass and radius with $\approx 13\%$ and $\approx 5\%$ precision, respectively. Notably, TOI-6651b is the densest sub-Saturn detected with TESS and the third densest among all known sub-Saturns to date. The discovery supports the theory of a positive correlation between planet mass and host star metallicity. TOI-6651b is estimated to have a core mass of $\approx 53\ M_\oplus$, composed predominantly of dense materials like rock and iron, which make around 87\% of the planet's total mass. The remaining mass fraction, with  $f_{env} \approx 0.13$, consists of a low-density H/He envelope. The formation history of TOI-6651b is unusual, possibly involving processes such as dynamical instabilities leading to mergers, tidal heating following high-eccentricity migration, and envelope accretion in a gas-depleted disk. Located at the edge of the Neptunian desert, TOI-6651b will be helpful in understanding the factors that define desert boundaries. Further observational and theoretical studies of dense sub-Saturns like TOI-6651b can provide deeper insights into their formation and evolution.


\begin{acknowledgements}
We extend our gratitude for the generous support provided by PRL-DOS (Department of Space, Government of India), as well as the Director, PRL. Their support has been instrumental in funding the PARAS-2 spectrograph for our exoplanet discovery project, as well as facilitating research grants for SB, AK, and SND. B.S.S. and I.A.S. acknowledge the support of M.V. Lomonosov Moscow State University Program of Development. We express our gratitude to all the Mount Abu Observatory staff for their invaluable assistance throughout the observations. SB also acknowledges Mr. Trinesh Sana for his assistance in executing \textit{Mathematica} code for the M-R diagram. This research has made use of the SIMBAD database and the VizieR catalog access tool, operated at CDS, Strasbourg, France. Additionally, this study utilized the Exoplanet Follow-up Observation Program (ExoFOP; DOI: 10.26134/ExoFOP5) website, managed by the California Institute of Technology under contract with the National Aeronautics and Space Administration as part of the Exoplanet Exploration Program. This paper includes data collected through the TESS mission, obtained from the MAST data archive at the Space Telescope Science Institute (STScI), and utilizes the TepCat and NASA Exoplanet Archive catalog. The authors also express gratitude to the anonymous referee for their numerous helpful suggestions, which significantly improved the quality of the paper.
\end{acknowledgements}


\bibliographystyle{aa}
\bibliography{ref}


\onecolumn

\begin{appendix}


\section{Additional table}

\begin{table}[h!]
\centering
\caption{Radial velocity measurements of TOI-6651 with PARAS-2$^*$.}
\label{tab:rv_table}
\begin{tabular}{ccccc}
\hline
\noalign{\smallskip}
BJD$_{TDB}$& Relative-RV & $\sigma$-RV & BIS & $\sigma$-BIS\\
Days & m s$^{-1}$ & m s$^{-1}$ & m s$^{-1}$ & m s$^{-1}$\\
\noalign{\smallskip}
\hline
\noalign{\smallskip}
2460266.212799 & 15.75 & 4.35 & -67.81 & 7.98 \\
2460269.227580 & -6.86 & 5.51 & -67.51 & 4.79 \\
2460270.209341 & 7.13 & 3.45 & -49.68 & 2.53 \\
2460271.202108 & 23.41 & 2.98 & -51.77 & 2.23 \\
2460271.252222 & 15.02 & 3.94 & -46.64 & 6.30 \\
2460272.235092 & -8.78 & 4.24 & -39.95 & 2.68 \\
2460273.137629 & -25.79 & 5.10 & -38.57 & 5.51 \\
2460273.198806 & -23.00 & 4.25 & -55.30 & 4.15 \\
2460279.123457 & -12.52 & 6.41 & -6.12 & 42.08 \\
2460286.258333 & 21.41 & 11.78 & -138.51 & 30.33 \\
2460287.188365 & -3.41 & 6.21 & -51.40 & 4.26 \\
2460288.197117 & -8.68 & 4.29 & -46.58 & 3.45 \\
2460290.159946 & -3.19 & 3.40 & -96.49 & 4.33 \\
2460293.174373 & -23.94 & 4.10 & -88.41 & 4.93 \\
2460295.161380 & 3.16 & 2.62 & -35.74 & 3.00 \\
2460302.168107 & 19.01 & 7.78 & -80.40 & 9.37 \\
2460302.217953 & 3.12 & 6.71 & -75.40 & 7.59 \\
2460303.170099 & -7.60 & 4.86 & -73.33 & 5.28 \\
2460305.156430 & -7.12 & 9.75 & -124.53 & 13.82 \\
2460306.157528 & 30.02 & 13.50 & -85.25 & 18.15 \\
2460307.106292 & 16.84 & 6.41 & -131.32 & 8.65 \\
2460307.159019 & 25.43 & 8.93 & -142.52 & 11.77 \\
2460308.101625 & -5.80 & 5.65 & -53.10 & 5.62 \\
2460308.157731 & -5.27 & 6.17 & -74.07 & 5.94 \\
2460309.110417 & -17.65 & 6.27 & -104.30 & 7.86 \\
2460325.086847 & -12.80 & 7.46 & -214.01 & 11.18 \\
2460330.098986 & 5.17 & 7.31 & -187.85 & 15.04 \\
\noalign{\smallskip}
\hline
\noalign{\smallskip}
\multicolumn{5}{l}{\footnotesize{$^*$Exposure time for each spectra is 3600 sec.}}
\end{tabular}
\end{table}


\newpage

\section{Additional figure}

\begin{figure*}[h!]
\centering
\caption{Corner plot showing the covariances for the fitted parameters from EXOFASTv2, corresponding to the high-mass solution (most probable) after splitting the bimodal distribution. The inner and outer contours around the median value represent the 68\% and 95\% confidence intervals, respectively.}
\label{fig:corner_circ}
\includegraphics[width=0.90\paperwidth]{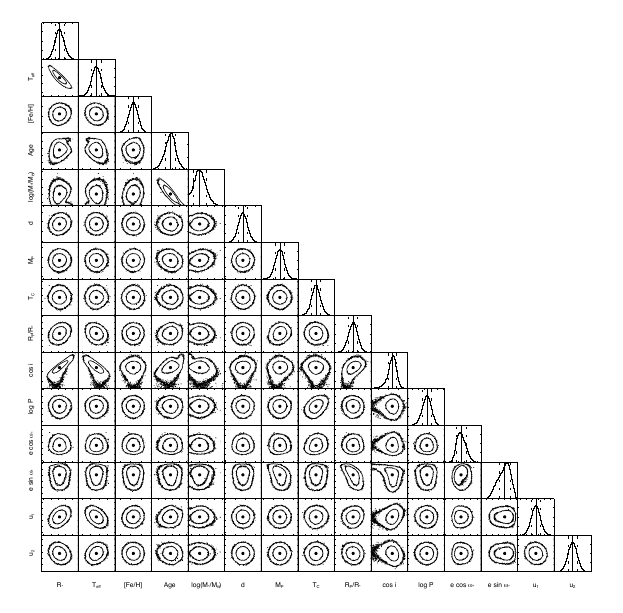}
\end{figure*}

\end{appendix}

\end{document}